\documentclass[12pt,reqno]{amsart}
\usepackage{graphicx}
\usepackage{amscd}
\usepackage{amsmath}
\usepackage{epsfig}
\usepackage{amsfonts}
\usepackage{amssymb}

\numberwithin{equation}{section}
\providecommand{\U}[1]{\protect\rule{.1in}{.1in}}
\providecommand{\U}[1]{\protect\rule{.1in}{.1in}}
\allowdisplaybreaks

\theoremstyle{plain}

\input{tcilatex}

\begin{document}
\title[The role of the Pauli-Luba\'{n}ski Vector]{The Role of the Pauli-Luba%
\'{n}ski Vector for the Dirac, Weyl, Proca, Maxwell, and Fierz-Pauli
Equations}
\author{Sergey I. Kryuchkov }
\address{School of Mathematical and Statistical Sciences, Arizona State
University, Tempe, AZ 85287--1804, U.S.A.}
\email{sergeykryuchkov@yahoo.com}
\author{Nathan A. Lanfear}
\address{School of Mathematical and Statistical Sciences, Arizona State
University, Tempe, AZ 85287--1804, U.S.A.}
\email{nlanfear@asu.edu}
\author{Sergei~K.~Suslov}
\address{School of Mathematical and Statistical Sciences, Arizona State
University, Tempe, AZ 85287--1804, U.S.A.}
\email{sks@asu.edu}
\urladdr{http://hahn.la.asu.edu/\symbol{126}suslov/index.html}
\date{January 22, 2016}
\subjclass{Primary 35Q41, 35Q61, 35L05; Secondary 35N05, 35Q75.}
\keywords{Poincar\'{e} group, Pauli-Luba\'{n}ski vector, Dirac's equation,
Weyl's equations for massless neutrinos, Proca and Maxwell's equations,
Fierz-Pauli equations, linearized Einstein equations.}

\begin{abstract}
We analyze basic relativistic wave equations for the classical fields, such
as Dirac's equation, Weyl's two-component equation for massless neutrinos,
and the Proca, Maxwell, and Fierz-Pauli equations, from the viewpoint of the
Pauli-Luba\'{n}ski vector and the Casimir operators of the Poincar\'{e}
group. In general, in this group-theoretical approach, the above wave
equations arise in certain overdetermined forms, which can be reduced to the
conventional ones by a Gaussian elimination. A~connection between the spin
of a particle/field and consistency of the corresponding overdetermined
system is emphasized in the massless case.
\end{abstract}

\maketitle

\section{Introduction}

All physically interesting representations of the proper orthochronous
inhomogeneous Lorentz group (known nowadays as the Poincar\'{e} group) were
classified by Wigner \cite{Wigner39} and, since then, this approach has been
utilized for the mathematical description of mass and spin of an elementary
particle. To this end, the Pauli-Luba\'{n}ski pseudo-vector is introduced,%
\begin{equation}
w_{\mu }=\frac{1}{2}e_{\mu \nu \sigma \tau }p^{\nu }M^{\sigma \tau },\qquad
p_{\mu }w^{\mu }=0,  \label{PauliLubanskiVector}
\end{equation}%
where $p_{\mu }$ is the relativistic linear momentum operator and $M^{\sigma
\tau }$ are the angular momentum operators, or generators of the proper
orthochronous Lorentz group, with summation over repeated indices. (We use
Einstein summation convention unless stated otherwise.) The mass and spin of
a particle are defined in terms of two quadratic invariants (Casimir
operators of the Poincar\'{e} group) as follows%
\begin{equation}
p^{2}=p_{\mu }p^{\mu }=m^{2},\qquad w^{2}=w_{\mu }w^{\mu }=-m^{2}s\left(
s+1\right) ,  \label{MassSpin}
\end{equation}%
when $m>0$ (see, for example, \cite{Bargmann54}, \cite{BargmannWigner48},
\cite{BarutRaczka80}, \cite{Bogolubovetal90}, \cite{Lub41}, \cite{Lub42-I},
\cite{Lub42-II}, \cite{RumerFetQFT}, \cite{Ryder}, \cite%
{ScheckQuantumPhysics}, \cite{Schweber61}, \cite{Sohnius85} and the
references therein; throughout the article we use the standard notations in
the Minkowski space-time $\left.
\mathbb{R}
\right. ^{4}$ and the natural units $c=\hslash =1).$

For the massless fields, when $m=0,$ one gets $w^{2}=p^{2}=pw=0,$ and the
Pauli-Luba\'{n}ski vector should be proportional to $p:\footnote{%
This assumption was made by Bargmann and Wigner \cite{BargmannWigner48} for
the massless limit of the spinor wave equation for particles with an
arbitrary integer or half-integer spin proposed by Dirac \cite{Dirac36} (see
also \cite{Fierz39}, \cite{Fierz40}, \cite{Kramersetal41}, \cite{Pauli41}
and the references therein). The pseudo-vector (\ref{PauliLubanskiVector})
was introduced, in a slightly different notation, by Eqs.~(4.a)--(4.b) of
Ref.~\cite{BargmannWigner48}.}$%
\begin{equation}
w_{\mu }=\lambda p_{\mu }  \label{WrongHelicity}
\end{equation}%
(acting on common eigenstates \cite{Naber12}, \cite{Ryder}). The number $%
\lambda $ is sometimes called the helicity of the representation and the
value $s=\left\vert \lambda \right\vert $ is called the spin of a particle
with zero mass \cite{Bogolubovetal90}, \cite{Ryder}, \cite%
{ScheckQuantumPhysics}, \cite{Schweber61}. Although the concept of helicity
is discussed in most textbooks on quantum field theory, a practical
implementation of this definition of the spin of a massless particle
deserves a certain clarification. As is shown in our previous article \cite%
{KrLanSus150}, in the case of the electromagnetic field in vacuum, the
constant $\lambda $ in the latter equation is fixed, otherwise violating the
classical Maxwell equations. Thus, for the photon, or a harmonic circular
classical electromagnetic wave\footnote{%
Multiple meanings of the word \textquotedblleft photon\textquotedblright\
are analyzed in \cite{Klyshko94}.}, the latter equation allows one to
introduce the field equations and spin, but not the helicity, when a certain
choice of eigenstates is required. A similar situation occurs in the case of
Weyl's equation for massless neutrinos. (In Ref.~\cite{KrLanSus150}, we do
not discuss the equation for a graviton, another massless spin-2 particle;
it will be analyzed elsewhere; cf. \cite{Khriplovich}, \cite{MisThoWhe},
\cite{MukhWin07}, \cite{Pauli} and our discussion in section~7.)

The theory of relativistic-invariant wave equations is studied, from
different perspectives, in numerous classical accounts \cite%
{BargmannWigner48}, \cite{BarutRaczka80}, \cite{Bhabha45}, \cite{Dirac28},
\cite{Dirac36}, \cite{Fierz39}, \cite{Fierz40}, \cite{Gelfandetal}, \cite%
{Kramersetal41}, \cite{Majorana32}, \cite{Pauli41}, \cite{Proca36}, \cite%
{Weyl} (see also \cite{Bia:Bia75}, \cite{Lub41}, \cite{Lub42-I}, \cite%
{Lub42-II} and the references therein). Nonetheless, in our opinion, the
importance of the Pauli-Luba\'{n}ski vector for conventional relativistic
equations, which allows one to derive all of them directly from the
postulated transformation law of the corresponding classical field in pure
group-theoretical terms, is not fully appreciated. In this article, we would
like to start from Dirac's relativistic electron, or any free relativistic
particle with a nonzero mass and spin $1/2,$ which can be described by a
bispinor wave function. Our analysis shows that an analog of the linear
operator relation (\ref{WrongHelicity}) takes the form,%
\begin{equation}
w_{\mu }=\frac{1}{2}\left( p_{\mu }+m\gamma _{\mu }\right) \gamma _{5},
\label{PauliLubanskiDirac}
\end{equation}%
provided that the Dirac equation, $\left( \gamma ^{\mu }p_{\mu }-m\right)
\psi =0,$ holds, when the corresponding overdetermined system of equations
is consistent. This automatically implies that $s=1/2,$ in the covariant
form, by definition (\ref{MassSpin}). (We were not able to find the operator
relation (\ref{PauliLubanskiDirac}) in the extensive literature on Dirac's
equation.)

In the rest of the article, a similar program is utilized, in a systematic
way and from first principles, for other familiar relativistic wave
equations. Once again, we postulate the transformation law of the field in
question (a law of nature) and, with the help of the corresponding Lorentz
generators, evaluate the action of the Pauli-Luba\'{n}ski vector on the
field in order to compute, eventually, not one, but both Casimir operators (%
\ref{MassSpin}). If a linear relation, similar to (\ref{WrongHelicity}) or (%
\ref{PauliLubanskiDirac}), does exist, one obtains an overdetermined PDE
system, which can be reduced to the corresponding relativistic wave equation
by a matrix version of Gaussian elimination \cite{Gantmacher1}. We show that
this approach allows one to derive equations of motion for the most useful
classical fields, including the Weyl, Proca, Fierz-Pauli, and Maxwell
equations in vacuum, as a statement of consistency for the original
overdetermined systems. At the moment, we shall not discuss relativistic
wave equations for particles with an arbitrary spin, such as the
Bargmann-Wigner equations, Majorana equations, and/or the (first order)
Duffin-Kemmer equations which also describe spin-0 and spin-1 fields (see,
for example, \cite{BarutRaczka80}, \cite{BogolubovShirkov}, \cite{Greiner}
and \cite{SchweberBetheHoffmann} for more details; the case of the
Klein-Gordon equation, or the relativistic Schr\"{o}dinger equation \cite%
{Schiff}, is, of course, obvious).

We have also entirely concentrated on four dimensions. A more general
group-theoretical approach to the relativistic wave equations, which allows
one to include higher dimensions and spins, is formulated in \cite%
{BarutRaczka80} with the help of induced representations of the semi-direct
products of separable, locally compact groups. Spinors in arbitrary
dimensions are also discussed in \cite{Sohnius85}.

\section{Dirac equation}

In this section, for the reader's convenience, we summarize basic facts
about Dirac's equation and then discuss its relation with the Pauli-Luba\'{n}%
ski vector (\ref{PauliLubanskiDirac}). To this end, a familiar bispinor
representation of the proper orthochronous Lorentz group $SO_{+}\left(
1,3\right) $ is used.

\subsection{Gamma matrices, bispinors, and transformation laws}

We shall use the following Dirac matrices: $\gamma^{\mu}=\left( \gamma ^{0},%
\boldsymbol{\gamma}\right) ,$ $\gamma_{\mu}=g_{\mu\nu}\gamma^{\nu}=$ $\left(
\gamma^{0},-\boldsymbol{\gamma}\right) ,$ and $\gamma_{5}=-\gamma^{5}=i%
\gamma^{0}\gamma^{1}\gamma^{2}\gamma^{3},$ where%
\begin{equation}
\boldsymbol{\gamma}=\left(
\begin{array}{cc}
0 & \boldsymbol{\sigma} \\
-\boldsymbol{\sigma} & 0%
\end{array}
\right) ,\qquad\gamma^{0}=\left(
\begin{array}{cc}
I & 0 \\
0 & -I%
\end{array}
\right) ,\qquad\gamma^{5}=\left(
\begin{array}{cc}
0 & I \\
I & 0%
\end{array}
\right)  \label{GammaDirac}
\end{equation}
and $\boldsymbol{\sigma}=\left( \sigma_{1},\sigma_{2},\sigma_{3}\right) $
are the standard $2\times2$ Pauli matrices \cite{Pauli36}, \cite{VMK}. The
familiar anticommutation relations,%
\begin{equation}
\gamma^{\mu}\gamma^{\nu}+\gamma^{\nu}\gamma^{\mu}=2g^{\mu\nu},\qquad
\gamma^{\mu}\gamma^{5}+\gamma^{5}\gamma^{\mu}=0\qquad\left( \mu
,\nu=0,1,2,3\right) ,  \label{AntiCommutator}
\end{equation}
hold. (Most of the results here will not depend on a particular choice of
gamma matrices, but it is always useful to have an example in mind.) The
four-vector notation, $x^{\mu}=\left( t,\mathbf{r}\right) ,$ $\partial_{\mu
}=\partial/\partial x^{\mu},$ and $\partial^{\alpha}=g^{\alpha\mu}\partial_{%
\mu}$ in natural units $c=\hslash=1$ with the standard metric, $%
g_{\mu\nu}=g^{\mu\nu}=\mathrm{diag}\left( 1,-1,-1,-1\right) ,$ in the
Minkowski space-time $\left.
\mathbb{R}
\right. ^{4}$ are utilized throughout the article \cite{Ber:Lif:Pit}, \cite%
{BogolubovShirkov}, \cite{Bogolubovetal90}, \cite{MisThoWhe}, \cite%
{SchweberBetheHoffmann}.

In this notation, the transformation law of a bispinor wave function%
\footnote{%
The relativistic wave equation for a massive spin $1/2$ particle was
proposed by Dirac \cite{Dirac28}, when only tensor representations of the
Lorentz group were known. Thus, the problem of covariance of Dirac's
equation gave rise to a new class of representations of the Lorentz group,
namely, the spinor representations \cite{BarutRaczka80}.},%
\begin{equation}
\psi \left( x\right) =\left(
\begin{array}{c}
\psi _{1} \\
\psi _{2} \\
\psi _{3} \\
\psi _{4}%
\end{array}%
\right) \in \left.
\mathbb{C}
\right. ^{4},  \label{bispinor}
\end{equation}%
under a proper Lorentz transformation, is given by%
\begin{equation}
\psi ^{\prime }\left( x^{\prime }\right) =S_{\Lambda }\psi \left( x\right)
,\qquad x^{\prime }=\Lambda x,  \label{TransformationLaw}
\end{equation}%
together with the rule,
\begin{equation}
S_{\Lambda }^{-1}\gamma ^{\mu }S_{\Lambda }=\Lambda _{\ \nu }^{\mu }\gamma
^{\nu },  \label{TransformationLawGamma}
\end{equation}%
for the sake of covariance of the celebrated Dirac equation,%
\begin{equation}
i\gamma ^{\mu }\partial _{\mu }\psi -m\psi =0.  \label{DiracEquationGamma}
\end{equation}

As is well known, a general solution of the latter matrix equation has the
form%
\begin{align}
S=S_{\Lambda }& =\exp \left( -\frac{1}{4}\theta _{\mu \nu }\Sigma ^{\mu \nu
}\right) ,\quad \theta _{\mu \nu }=-\theta _{\nu \mu },  \label{MatrixS} \\
\Sigma ^{\mu \nu }& =\left( \gamma ^{\mu }\gamma ^{\nu }-\gamma ^{\nu
}\gamma ^{\mu }\right) /2  \notag
\end{align}%
(with summation over every two repeated indices; see, for example, \cite%
{ItzZub}, \cite{MoskalevQFT}) and, in turn,%
\begin{equation}
S_{\Lambda }^{-1}\Sigma ^{\mu \nu }S_{\Lambda }=\Lambda _{\ \sigma }^{\mu
}\Lambda _{\ \tau }^{\nu }\Sigma ^{\sigma \tau }.
\label{TransformationLawSigma}
\end{equation}%
In explicit form,%
\begin{align}
\Sigma ^{\mu \nu }& =\frac{1}{2}\left( \gamma ^{\mu }\gamma ^{\nu }-\gamma
^{\nu }\gamma ^{\mu }\right) =\left(
\begin{array}{cc}
0 & \alpha _{q}\medskip \\
-\alpha _{p} & -ie_{pqr}\Sigma _{r}%
\end{array}%
\right)  \label{Sigma} \\
& =\left(
\begin{array}{cccc}
0 & \alpha _{1} & \alpha _{2} & \alpha _{3} \\
-\alpha _{1} & 0 & -i\Sigma _{3} & i\Sigma _{2} \\
-\alpha _{2} & i\Sigma _{3} & 0 & -i\Sigma _{1} \\
-\alpha _{3} & -i\Sigma _{2} & i\Sigma _{1} & 0%
\end{array}%
\right) ,  \notag
\end{align}%
where, by definition,%
\begin{equation}
\boldsymbol{\Sigma }=\left(
\begin{array}{cc}
\boldsymbol{\sigma } & 0 \\
0 & \boldsymbol{\sigma }%
\end{array}%
\right) ,\qquad \boldsymbol{\alpha }=\left(
\begin{array}{cc}
0 & \boldsymbol{\sigma } \\
\boldsymbol{\sigma } & 0%
\end{array}%
\right) .  \label{SigmaAlpha}
\end{equation}%
Their familiar product identities,%
\begin{align}
& \Sigma _{p}\Sigma _{q}=ie_{pqr}\Sigma _{r}+\delta _{pq},\qquad \alpha
_{p}\alpha _{q}=ie_{pqr}\Sigma _{r}+\delta _{pq},  \label{alphaSigmaProducts}
\\
& \qquad \alpha _{p}\Sigma _{q}=\Sigma _{p}\alpha _{q}=ie_{pqr}\alpha
_{r}+\delta _{pq}\gamma ^{5},  \notag
\end{align}%
hold.

Setting $\boldsymbol{n}=\left\{ \mathbf{e}_{1},\mathbf{e}_{2},\mathbf{e}%
_{3}\right\} $ for each of the unit vectors in the directions of the
mutually orthogonal coordinate axes, one can write in compact form:
\begin{equation}
S_{R}=e^{i\theta \left( \boldsymbol{n\cdot \Sigma }\right) /2}=\cos \frac{%
\theta }{2}+i\left( \boldsymbol{n\cdot \Sigma }\right) \sin \frac{\theta }{2}%
,\qquad \left( \boldsymbol{n\cdot \Sigma }\right) ^{2}=I
\label{DiracRotations}
\end{equation}%
and%
\begin{equation}
S_{L}=e^{-\vartheta \left( \boldsymbol{n\cdot \alpha }\right) /2}=\cosh
\frac{\vartheta }{2}-\left( \boldsymbol{n\cdot \alpha }\right) \sinh \frac{%
\vartheta }{2},\qquad \left( \boldsymbol{n\cdot \alpha }\right) ^{2}=I
\label{DiracBoosts}
\end{equation}%
with $\tanh \vartheta =v,$ in the cases of rotations and boosts,
respectively \cite{BarutRaczka80}, \cite{MoskalevQFT}.

The important dual four-tensor identities,%
\begin{equation}
ie_{\mu\nu\sigma\tau}\Sigma^{\sigma\tau}=2\gamma_{5}\Sigma_{\mu\nu},\qquad
ie^{\mu\nu\sigma\tau}\gamma_{5}\Sigma_{\mu\nu}=2\Sigma^{\sigma\tau},
\label{DualGenerators}
\end{equation}
can be directly verified. Here, $e^{\mu\nu\sigma\tau}=-e_{\mu\nu\sigma\tau}$
and $e_{0123}=+1$ is the Levi-Civita symbol \cite{Bogolubovetal90}, \cite%
{Fock64}.

For the conjugate bispinor,%
\begin{equation}
\overline{\psi }\left( x\right) =\psi ^{\dag }\left( x\right) \gamma
^{0},\qquad \overline{\psi }^{\prime }\left( x^{\prime }\right) =\overline{%
\psi }\left( x\right) S_{\Lambda }^{-1},\qquad x^{\prime }=\Lambda x,
\label{bispinorDconjugate}
\end{equation}%
the Dirac equation (\ref{DiracEquationGamma}) takes the form%
\begin{equation}
i\partial _{\mu }\overline{\psi }\gamma ^{\mu }+m\overline{\psi }=0.
\label{DiracEquationConjugate}
\end{equation}%
(For more details see classical accounts \cite{Akh:Ber}, \cite{Ber:Lif:Pit},
\cite{BogolubovShirkov}, \cite{Feynman}, \cite{ItzZub}, \cite{MoskalevQFT},
\cite{PeskinSchroeder}, \cite{Ryder}, \cite{Schweber61}, \cite%
{SchweberBetheHoffmann}, \cite{Wein}.)

\noindent \textbf{Examples.} In particular, for the boost in the plane $%
\left( x^{0},x^{1}\right) ,$ when%
\begin{equation}
S_{L}=e^{-\left( \vartheta /2\right) \Sigma ^{01}}=e^{-\left( \vartheta
/2\right) \alpha _{1}}=\cosh \frac{\vartheta }{2}-\alpha _{1}\sinh \frac{%
\vartheta }{2}  \label{BoostDirac01}
\end{equation}%
with $\tanh \vartheta =v,$ one can easily verify by matrix multiplication
that%
\begin{equation}
e^{\left( \vartheta /2\right) \alpha _{1}}\left(
\begin{array}{c}
\gamma ^{0} \\
\gamma ^{1} \\
\gamma ^{2} \\
\gamma ^{3}%
\end{array}%
\right) e^{-\left( \vartheta /2\right) \alpha _{1}}=\left(
\begin{array}{cccc}
\cosh \vartheta & -\sinh \vartheta & 0 & 0 \\
-\sinh \vartheta & \cosh \vartheta & 0 & 0 \\
0 & 0 & 1 & 0 \\
0 & 0 & 0 & 1%
\end{array}%
\right) \left(
\begin{array}{c}
\gamma ^{0} \\
\gamma ^{1} \\
\gamma ^{2} \\
\gamma ^{3}%
\end{array}%
\right) .
\end{equation}%
In a similar fashion, for the rotation in the plane $\left(
x^{1},x^{2}\right) :$%
\begin{equation}
S_{R}=e^{-\left( \theta /2\right) \Sigma ^{12}}=e^{i\left( \theta /2\right)
\Sigma _{3}}=\cos \frac{\theta }{2}+i\Sigma _{3}\sin \frac{\theta }{2}
\label{RotationDirac12}
\end{equation}%
and%
\begin{equation}
e^{-i\left( \theta /2\right) \Sigma _{3}}\left(
\begin{array}{c}
\gamma ^{0} \\
\gamma ^{1} \\
\gamma ^{2} \\
\gamma ^{3}%
\end{array}%
\right) e^{i\left( \theta /2\right) \Sigma _{3}}=\left(
\begin{array}{cccc}
1 & 0 & 0 & 0 \\
0 & \cos \theta & \sin \theta & 0 \\
0 & -\sin \theta & \cos \theta & 0 \\
0 & 0 & 0 & 1%
\end{array}%
\right) \left(
\begin{array}{c}
\gamma ^{0} \\
\gamma ^{1} \\
\gamma ^{2} \\
\gamma ^{3}%
\end{array}%
\right) .
\end{equation}%
(See \cite{BarutRaczka80}, \cite{MoskalevQFT} for further details.)

\subsection{Generators and commutators}

In the fundamental representation of the proper orthochronous Lorentz group,
we shall choose the following six $4\times 4$ real-valued matrices $(\alpha
,\beta =0,1,2,3$ are fixed with no summation):%
\begin{align}
\Lambda \left( \theta _{\alpha \beta }\right) & =\exp \left( -\theta
_{\alpha \beta }\ m^{\alpha \beta }\right) ,\qquad m^{\alpha \beta
}=-m^{\beta \alpha },  \label{OneParameter} \\
\left( m^{\alpha \beta }\right) _{\nu }^{\mu }& =g^{\alpha \mu }\delta _{\nu
}^{\beta }-g^{\beta \mu }\delta _{\nu }^{\alpha }  \notag
\end{align}%
for the corresponding one-parameter subgroups of rotations and boosts \cite%
{Bogolubovetal90}, \cite{KrLanSus150}, \cite{MisThoWhe}, \cite{Schweber61}.
Then, differentiation of a particular expression (for the corresponding
tensor operator \cite{BarutRaczka80}),%
\begin{equation}
e^{\theta _{\alpha \beta }\Sigma ^{\alpha \beta }/2}\gamma ^{\mu }e^{-\theta
_{\alpha \beta }\Sigma ^{\alpha \beta }/2}=\left( e^{-\theta _{\alpha \beta
}m^{\alpha \beta }}\right) _{\nu }^{\mu }\gamma ^{\nu },
\label{OneparameterGamma}
\end{equation}%
at $\theta _{\alpha \beta }=0$ results in%
\begin{equation}
\left[ \Sigma ^{\alpha \beta },\ \gamma ^{\mu }\right] :=\Sigma ^{\alpha
\beta }\gamma ^{\mu }-\gamma ^{\mu }\Sigma ^{\alpha \beta }=2\left( g^{\beta
\mu }\gamma ^{\alpha }-g^{\alpha \mu }\gamma ^{\beta }\right) ,
\label{CommutatorSigmaGamma}
\end{equation}%
which can be independently verified with the help of (\ref{AntiCommutator}).

In a similar fashion, the action of four-angular momentum operators\footnote{%
We follow \cite{KrLanSus150}. Traditionally, $M^{\alpha \beta }\rightarrow
-iM^{\alpha \beta };$ see, for example, \cite{Bogolubovetal90}.},%
\begin{equation}
M^{\alpha \beta }=x^{\beta }\partial ^{\alpha }-x^{\alpha }\partial ^{\beta
},\qquad \partial ^{\alpha }=g^{\alpha \kappa }\partial _{\kappa },
\label{DiffGenerators}
\end{equation}%
on Dirac's bispinors (\ref{bispinor}) can be derived directly from the
transformation law as follows%
\begin{align}
M^{\alpha \beta }\psi & :=-\left. \left[ \frac{d}{d\theta _{\alpha \beta }}%
\psi ^{\prime }\left( \Lambda _{\ \nu }^{\mu }\left( \theta _{\alpha \beta
}\right) x^{\nu }\right) \right] \right\vert _{\theta _{\alpha \beta }=0}
\label{Generators} \\
& =-\left. \left( \frac{d}{d\theta _{\alpha \beta }}e^{-\theta _{\alpha
\beta }\Sigma ^{\alpha \beta }/2}\right) \right\vert _{\theta _{\alpha \beta
}=0}\psi \left( x\right) =\frac{1}{2}\Sigma ^{\alpha \beta }\psi  \notag
\end{align}%
(see also \cite{ItzZub} for a slightly different derivation). A familiar
commutator,%
\begin{equation}
\left[ \Sigma ^{\alpha \beta },\ \Sigma ^{\sigma \tau }\right] =2\left(
g^{\beta \sigma }\Sigma ^{\alpha \tau }-g^{\beta \tau }\Sigma ^{\alpha
\sigma }+g^{\alpha \tau }\Sigma ^{\beta \sigma }-g^{\alpha \sigma }\Sigma
^{\beta \tau }\right) ,  \label{CommutatorSigma}
\end{equation}%
can be readily verified with the help of (\ref{CommutatorSigmaGamma}) and/or
independently derived from (\ref{TransformationLawSigma}). These results are
independent of our choice of the gamma matrices representation.

\subsection{Balance conditions and energy-momentum tensors}

We shall use a familiar notation for the partial derivatives \cite%
{Bogolubovetal90},
\begin{equation}
\mathcal{D}^{p}\psi\left( x\right) :=\frac{\partial^{p_{0}+p_{1}+p_{2}+p_{3}}%
}{\partial x_{0}^{p_{0}}\partial x_{1}^{p_{1}}\partial x_{2}^{p_{2}}\partial
x_{3}^{p_{3}}}\psi\left( x_{0},x_{1},x_{2},x_{3}\right) ,\qquad\mathcal{D}%
^{0}\psi\left( x\right) :=\psi\left( x\right)  \label{PartialDerivatives}
\end{equation}
where $p=\left( p_{0},p_{1},p_{2},p_{3}\right) $ is an ordered set of
non-negative integers $p_{\mu}\geq0.$ It follows from the Dirac equations (%
\ref{DiracEquationGamma}) and (\ref{DiracEquationConjugate}) that
\begin{equation}
\partial_{\mu}\left[ \left( \mathcal{D}^{p}\overline{\psi}\left( x\right)
\right) \gamma^{\mu}\left( \mathcal{D}^{q}\psi\left( x\right) \right) \right]
=0,  \label{BalanceMulti}
\end{equation}
or, for a finite multi-sum,%
\begin{equation}
\partial_{\mu}\left[ \sum_{p,q}c_{p,q}\mathcal{D}^{p}\overline{\psi}%
\gamma^{\mu}\mathcal{D}^{q}\psi\right] =0,  \label{BalanceSum}
\end{equation}
which can be thought of as a \textquotedblleft master\textquotedblright\
differential balance condition set.

Indeed, in view of $\mathcal{D}^{r}\partial_{\mu}=\partial_{\mu}\mathcal{D}%
^{r},$ one gets%
\begin{equation}
i\gamma^{\mu}\partial_{\mu}\left( \mathcal{D}^{q}\psi\right) =m\left(
\mathcal{D}^{q}\psi\right) ,\qquad i\partial_{\mu}\left( \mathcal{D}^{p}%
\overline{\psi}\right) \gamma^{\mu}=-m\left( \mathcal{D}^{p}\overline{\psi}%
\right) .
\end{equation}
Let us multiply the first (second) equation by $\mathcal{D}^{p}\overline{%
\psi }$ ($\mathcal{D}^{q}\psi$) from the left (right) and add the results.
Then%
\begin{align*}
i\partial_{\mu}\left( \mathcal{D}^{p}\overline{\psi}\gamma^{\mu}\mathcal{D}%
^{q}\psi\right) & =i\partial_{\mu}\left( \mathcal{D}^{p}\overline{\psi}%
\right) \gamma^{\mu}\mathcal{D}^{q}\psi+i\mathcal{D}^{p}\overline{\psi}%
\gamma^{\mu}\partial_{\mu}\left( \mathcal{D}^{q}\psi\right) \\
& =m\left( -\mathcal{D}^{p}\overline{\psi}\mathcal{D}^{q}\psi+\mathcal{D}^{p}%
\overline{\psi}\mathcal{D}^{q}\psi\right) \equiv0.
\end{align*}

Among important special cases of (\ref{BalanceMulti}) are the following
important identities:%
\begin{equation}
\partial_{\mu}j^{\mu}\left( x\right) =0,\qquad j^{\mu}\left( x\right) =%
\overline{\psi}\left( x\right) \gamma^{\mu}\psi\left( x\right) ,
\label{DiracChargeConservation}
\end{equation}
corresponding to the total charge conservation, and%
\begin{equation}
\partial_{\mu}\left[ \left. \overline{\psi}\left( x\right) \gamma^{\mu
}\partial_{\nu}\psi\left( x\right) \right. \right] =0,\qquad\partial_{\mu }%
\left[ \left. \partial_{\nu}\overline{\psi}\left( x\right) \gamma^{\mu
}\psi\left( x\right) \right. \right] =0.
\end{equation}
Therefore, one can introduce the energy-momentum tensor, such that $%
\partial_{\mu}T_{\ \nu}^{\mu}\left( x\right) =0,$ in two different forms%
\begin{equation}
T_{\ \nu}^{\mu}:=i\overline{\psi}\gamma^{\mu}\partial_{\nu}\psi,\qquad T_{\
\mu}^{\mu}=m\overline{\psi}\psi  \label{EMTensor1Dirac}
\end{equation}
and/or%
\begin{equation}
T_{\ \nu}^{\mu}:=\frac{i}{2}\left[ \left. \overline{\psi}\gamma^{\mu}\left(
\partial_{\nu}\psi\right) -\left( \partial_{\nu}\overline{\psi}\right)
\gamma^{\mu}\psi\right. \right] ,\qquad T_{\ \mu}^{\mu}=m\overline{\psi}\psi.
\label{EMTensor2Dirac}
\end{equation}
As is well-known, all quantities of physical interest can be derived from
the energy-momentum tensor \cite{BogolubovShirkov}, \cite{Schweber61}.

\subsection{Variants of Dirac's equation}

In view of (\ref{DiracEquationGamma}) and (\ref{AntiCommutator}),%
\begin{align}
& i\gamma ^{\mu }\gamma ^{\nu }\partial _{\nu }\psi =m\gamma ^{\mu }\psi ,
\label{GammatoSigma} \\
& \gamma ^{\mu }\gamma ^{\nu }=\frac{1}{2}\left( \gamma ^{\mu }\gamma ^{\nu
}-\gamma ^{\nu }\gamma ^{\mu }\right) +\frac{1}{2}\left( \gamma ^{\mu
}\gamma ^{\nu }+\gamma ^{\nu }\gamma ^{\mu }\right) =\Sigma ^{\mu \nu
}+g^{\mu \nu },  \notag
\end{align}%
and, as a result, we obtain an overdetermined but very convenient form of
Dirac's system:%
\begin{equation}
i\left( \Sigma ^{\mu \nu }+g^{\mu \nu }\right) \partial _{\nu }\psi =m\gamma
^{\mu }\psi .  \label{DiracEquationSigma}
\end{equation}%
On the one hand, in a $3D$ \textquotedblleft vector\textquotedblright\ form,%
\begin{equation}
i\partial _{0}\psi =\left( -i\boldsymbol{\alpha }\cdot \nabla +m\beta
\right) \psi ,\quad \beta =\gamma ^{0}\quad \left( \hslash =c=1,\quad
\partial _{0}=\partial /\partial t\right)  \label{DiracAlphaBeta}
\end{equation}%
and%
\begin{equation}
-i\partial _{0}\boldsymbol{\alpha }\psi +\left( \nabla \times \boldsymbol{%
\Sigma }\right) \psi =\left( i\nabla +m\boldsymbol{\gamma }\right) \psi .
\label{DiracAlphaSigma}
\end{equation}%
It is worth noting that the latter vectorial equation in our overdetermined
system (\ref{DiracAlphaBeta})--(\ref{DiracAlphaSigma}) can be obtained by
matrix multiplication, from the first one, in view of familiar relations:%
\begin{equation}
\alpha _{p}\alpha _{q}=ie_{pqr}\Sigma _{r}+\delta _{pq},\qquad \boldsymbol{%
\alpha }\beta =-\boldsymbol{\gamma }.  \label{AlphaProduct}
\end{equation}

On the other hand, by letting $p_{\mu }=i\partial _{\mu },$ one gets%
\begin{equation}
\left( \Sigma ^{\mu \nu }+g^{\mu \nu }\right) p_{\nu }\psi =m\gamma ^{\mu
}\psi  \label{DiracEquationP}
\end{equation}%
and applying the momentum operator $p_{\mu }$ to the both sides:
\begin{equation}
p^{2}\psi =\left( \Sigma ^{\mu \nu }+g^{\mu \nu }\right) p_{\mu }p_{\nu
}\psi =m\gamma ^{\mu }p_{\mu }\psi ,\qquad  \label{dAlembertP}
\end{equation}%
in view of $\Sigma ^{\mu \nu }p_{\mu }p_{\nu }=0.$ If $p^{2}\psi =m^{2}\psi
, $ we derive, once again, that $\gamma ^{\mu }p_{\mu }\psi =m\psi .$
Therefore both forms of the Dirac system, (\ref{DiracEquationGamma}) and (%
\ref{DiracEquationSigma}), are equivalent and every component of the
bispinor (\ref{bispinor}) does satisfy the d'Alembert equation,%
\begin{equation}
\left( \partial ^{\mu }\partial _{\mu }+m^{2}\right) \psi =\left( \partial
_{tt}^{2}-\Delta +m^{2}\right) \psi =\left( \square +m^{2}\right) \psi =0,
\label{dAlembertD}
\end{equation}%
as required by (\ref{MassSpin}).

In a similar fashion, for the conjugate bispinor (\ref{bispinorDconjugate})
one can obtain%
\begin{equation}
i\partial_{\nu}\overline{\psi}\left( \Sigma^{\mu\nu}-g^{\mu\nu}\right) =m%
\overline{\psi}\gamma^{\mu}  \label{DiracEquationSigmaConjugate}
\end{equation}
and our equations (\ref{DiracEquationSigma}) and (\ref%
{DiracEquationSigmaConjugate}) results in the following balance relation:%
\begin{equation}
i\partial_{\nu}\left( \overline{\psi}\Sigma^{\mu\nu}\psi\right) +i\left[
\left. \overline{\psi}\partial^{\mu}\psi-\left( \partial^{\mu}\overline {\psi%
}\right) \psi\right. \right] =2m\overline{\psi}\gamma^{\mu}\psi,
\label{balanceDirac}
\end{equation}
which, in turn, implies a familiar conservation law, $\partial_{\mu}j^{\mu
}\left( x\right) =0,$ for the four-current, $j^{\mu}\left( x\right) =%
\overline{\psi}\left( x\right) \gamma^{\mu}\psi\left( x\right) ,$ in view of
$\partial_{\mu}\partial_{\nu}\left( \overline{\psi}\Sigma^{\mu\nu}\psi%
\right) \equiv0$ and $\partial_{\mu}\left[ \left. \overline{\psi }%
\partial^{\mu}\psi-\left( \partial^{\mu}\overline{\psi}\right) \psi\right. %
\right] \equiv0.$ (The latter equation gives a differential balance
condition on its own.)

The following identity,%
\begin{equation}
i\partial _{\nu }\left( \overline{\psi }\Sigma ^{\mu \nu }\gamma ^{\lambda
}\psi \right) =2i\overline{\psi }\gamma ^{\mu }\partial ^{\lambda }\psi -i%
\left[ \left. \overline{\psi }\gamma ^{\lambda }\partial ^{\mu }\psi -\left(
\partial ^{\mu }\overline{\psi }\right) \gamma ^{\lambda }\psi \right. %
\right] ,  \label{balanceEMDirac}
\end{equation}%
can be obtained with the help of (\ref{DiracEquationSigma}), (\ref%
{DiracEquationSigmaConjugate}) and (\ref{CommutatorSigmaGamma}). The latter
shows how the difference between two forms of the energy-momentum tensor (%
\ref{EMTensor1Dirac}) and (\ref{EMTensor2Dirac}) can be written as the
four-divergence of a given tensor. Moreover, one can write%
\begin{equation}
T^{\mu \nu }:=i\overline{\psi }\gamma ^{\mu }\partial ^{\nu }\psi =mg^{\mu
\nu }\overline{\psi }\psi +m\overline{\psi }\Sigma ^{\mu \nu }\psi -i%
\overline{\psi }\gamma ^{\mu }\Sigma ^{\nu \lambda }\partial _{\lambda }\psi
\label{EMDirac1}
\end{equation}%
and%
\begin{align}
T^{\mu \nu }& :=\frac{i}{2}\left[ \left. \overline{\psi }\gamma ^{\mu
}\left( \partial ^{\nu }\psi \right) -\left( \partial ^{\nu }\overline{\psi }%
\right) \gamma ^{\mu }\psi \right. \right]  \label{EMDirac2} \\
& =mg^{\mu \nu }\overline{\psi }\psi -\frac{i}{2}\left[ \left. \overline{%
\psi }\gamma ^{\mu }\Sigma ^{\nu \lambda }\partial _{\lambda }\psi +\left(
\partial _{\lambda }\overline{\psi }\right) \Sigma ^{\nu \lambda }\gamma
^{\mu }\psi \right. \right]  \notag
\end{align}%
in view of (\ref{DiracEquationSigma}) and (\ref{DiracEquationSigmaConjugate}%
).

\subsection{Covariance and transformation of generators}

The relativistic invariance of the Dirac equation is a fundamental
consequence of (\ref{TransformationLaw})--(\ref{DiracEquationGamma}); see
for example, \cite{Ber:Lif:Pit}, \cite{BogolubovShirkov}, \cite{ItzZub},
\cite{MoskalevQFT}, \cite{SchweberBetheHoffmann} and \cite{Schweber61}.
Covariance of system (\ref{DiracEquationSigma}) can be derived, in a similar
fashion, by invoking (\ref{TransformationLawSigma}). The details are left to
the reader (see also section~5.3).

It is worth noting that from the four-tensor character of $\Sigma^{\mu\nu},$
in (\ref{TransformationLawSigma}), follow the transformation laws for the
generators of rotations and boosts. Let us assume that the velocity vector $%
\boldsymbol{v},$ for going over to a moving frame of reference, has the
direction of one of the coordinate axes, say $\left\{ \mathbf{e}_{a}\right\}
_{a=1,2,3}.$ Consider also \textquotedblleft orthogonal
decompositions\textquotedblright, $\boldsymbol{\Sigma}=\boldsymbol{\Sigma }%
_{\parallel}+\boldsymbol{\Sigma}_{\perp}$ and $\boldsymbol{\alpha }=%
\boldsymbol{\alpha}_{\parallel}+\boldsymbol{\alpha}_{\perp},$ in the
corresponding parallel and perpendicular directions, respectively. Then,
under the Lorentz transformation,%
\begin{equation}
S_{\Lambda}^{-1}\boldsymbol{\Sigma}_{\parallel}S_{\Lambda}=\boldsymbol{%
\Sigma }_{\parallel},\qquad S_{\Lambda}^{-1}\boldsymbol{\alpha}%
_{\parallel}S_{\Lambda}=\boldsymbol{\alpha}_{\parallel}
\label{GeneratorsParallel}
\end{equation}
and%
\begin{equation}
S_{\Lambda}^{-1}\boldsymbol{\Sigma}_{\perp}S_{\Lambda}=\frac {\boldsymbol{%
\Sigma}_{\perp}-i\left( \boldsymbol{v}\times\boldsymbol{\alpha }\right) }{%
\sqrt{1-v^{2}}},\qquad S_{\Lambda}^{-1}\boldsymbol{\alpha}_{\perp
}S_{\Lambda}=\frac{\boldsymbol{\alpha}_{\perp}-i\left( \boldsymbol{v}\times%
\boldsymbol{\Sigma}\right) }{\sqrt{1-v^{2}}},
\label{GeneratorsPerpendicular}
\end{equation}
by analogy with the transformations of electromagnetic fields\ in classical
electrodynamics \cite{KrLanSus150}, \cite{KrLanSu15}, \cite%
{SchweberBetheHoffmann}, \cite{ToptyginI}.

The corresponding invariants \cite{MoskalevQFT} are given by%
\begin{equation}
I_{1}=\Sigma _{\mu \nu }\Sigma ^{\mu \nu }=-2\left( \boldsymbol{\alpha }^{2}+%
\boldsymbol{\Sigma }^{2}\right) =-12  \label{InvariantOne}
\end{equation}%
and%
\begin{equation}
I_{2}=e_{\mu \nu \sigma \tau }\Sigma ^{\mu \nu }\Sigma ^{\sigma \tau }=-8i%
\boldsymbol{\alpha }\cdot \boldsymbol{\Sigma }=-2i\Sigma ^{\mu \nu }\Sigma
_{\mu \nu }\gamma _{5}=-2iI_{1}\gamma _{5}=24i\gamma _{5},
\label{InvariantTwo}
\end{equation}%
in view of the first identity (\ref{DualGenerators}). The invariant nature
of the helicity and relativistic rotation of the particle spin \cite%
{MoskalevQFT} can be naturally explained from these transformations.

\noindent \textbf{Example.} If $\boldsymbol{v}=v\mathbf{e}_{1},$ we set $%
\boldsymbol{\Sigma }_{\parallel }=\Sigma _{1},$ $\boldsymbol{\Sigma }_{\perp
}=\left\{ \Sigma _{2},\Sigma _{3}\right\} $ and $\boldsymbol{\alpha }%
_{\parallel }=\alpha _{1},$ $\boldsymbol{\alpha }_{\perp }=\left\{ \alpha
_{2},\alpha _{3}\right\} $ for the boost $S_{L}$ given by (\ref{BoostDirac01}%
). By the transformation law (\ref{GeneratorsParallel}), one gets $%
S^{-1}\Sigma _{1}S=\Sigma _{1},$ $S^{-1}\alpha _{1}S=\alpha _{1},$ which is
evident. In view of (\ref{GeneratorsPerpendicular}), the following matrix
identities,%
\begin{equation}
S^{-1}\Sigma _{2}S=\frac{\Sigma _{2}+iv\alpha _{3}}{\sqrt{1-v^{2}}},\qquad
S^{-1}\Sigma _{3}S=\frac{\Sigma _{3}-iv\alpha _{2}}{\sqrt{1-v^{2}}}
\end{equation}%
and%
\begin{equation}
S^{-1}\alpha _{2}S=\frac{\alpha _{2}+iv\Sigma _{3}}{\sqrt{1-v^{2}}},\qquad
S^{-1}\alpha _{3}S=\frac{\alpha _{3}-iv\Sigma _{2}}{\sqrt{1-v^{2}}},
\end{equation}%
hold. Indeed, in the first relation,%
\begin{align*}
S^{-1}\Sigma _{2}S& =\left( \cosh \frac{\vartheta }{2}+\alpha _{1}\sinh
\frac{\vartheta }{2}\right) \Sigma _{2}\left( \cosh \frac{\vartheta }{2}%
-\alpha _{1}\sinh \frac{\vartheta }{2}\right) \\
& =\Sigma _{2}\cosh ^{2}\frac{\vartheta }{2}+\left( \alpha _{1}\Sigma
_{2}-\Sigma _{2}\alpha _{1}\right) \cosh \frac{\vartheta }{2}\sinh \frac{%
\vartheta }{2}-\alpha _{1}\Sigma _{2}\alpha _{1}\sinh ^{2}\frac{\vartheta }{2%
} \\
& =\Sigma _{2}\cosh \vartheta +i\alpha _{3}\sinh \vartheta =\frac{\Sigma
_{2}+iv\alpha _{3}}{\sqrt{1-v^{2}}},
\end{align*}%
provided $\tanh \vartheta =v.$ Verifications of the remaining identities are
similar.

\subsection{Hamiltonian and energy balance}

The energy-momentum (density) four-vector is given by $T^{0\nu }$ and, in
view of (\ref{EMDirac1}) and (\ref{DiracAlphaBeta}), one gets%
\begin{align}
T^{00}& =i\overline{\psi }\gamma ^{0}\partial ^{0}\psi =m\overline{\psi }%
\psi -i\overline{\psi }\gamma ^{0}\Sigma ^{0p}\partial _{p}\psi
\label{DiracEnergy} \\
& =\psi ^{\dag }\left( -i\boldsymbol{\alpha }\cdot \nabla +m\beta \right)
\psi =\psi ^{\dag }H\psi .  \notag
\end{align}%
Here,%
\begin{equation}
i\partial _{0}\psi =H\psi ,\qquad H=-i\boldsymbol{\alpha }\cdot \nabla
+m\beta ,  \label{DiracHamiltonian}
\end{equation}%
which presents a familiar Hamiltonian form of the Dirac equation. The
differential balance equation take the form%
\begin{equation}
\partial _{0}\left( \psi ^{\dag }H\psi \right) +\func{div}\left( \psi ^{\dag
}\boldsymbol{\alpha }H\psi \right) =0,  \label{EnergyBalance}
\end{equation}%
where $\boldsymbol{\alpha }H=\nabla \times \boldsymbol{\Sigma }-i\nabla -m%
\boldsymbol{\gamma }$ in view of (\ref{AlphaProduct}).

\subsection{The Pauli-Luba\'{n}ski vector and Dirac's equation for a free
particle}

One can easily see that equation (\ref{DiracEquationSigma}) is related to
the Pauli-Luba\'{n}ski vector in view of the dual identities (\ref%
{DualGenerators}). Indeed,%
\begin{equation}
\Sigma^{\mu\nu}=\frac{i}{2}e^{\mu\nu\sigma\tau}\left(
\gamma_{5}\Sigma_{\sigma\tau}\right)
,\qquad\gamma_{5}\Sigma_{\sigma\tau}=\Sigma_{\sigma\tau}\gamma_{5},
\label{DiracSigmaDuality}
\end{equation}
and%
\begin{equation*}
\frac{i}{2}e^{\mu\nu\sigma\tau}\gamma_{5}\Sigma_{\sigma\tau}p_{\nu}\psi=-g^{%
\mu\nu}\left( p_{\nu}-m\gamma_{\nu}\right) \psi.
\end{equation*}
By \textquotedblleft index manipulations\textquotedblright,%
\begin{align}
& \frac{1}{2}e_{\mu\nu\sigma\tau}\left( i\partial^{\nu}\right) \left(
-i\Sigma^{\sigma\tau}\right) \psi=\frac{1}{2}e_{\mu\nu\sigma\tau}\partial^{%
\nu}\Sigma^{\sigma\tau}\psi \\
& \quad=\gamma_{5}\left( i\partial_{\mu}-m\gamma_{\mu}\right) \psi=\left(
i\partial_{\mu}+m\gamma_{\mu}\right) \gamma_{5}\psi  \notag
\end{align}
with the help of familiar properties of the gamma matrices, namely, $%
\gamma_{5}^{2}=I$ and $\gamma_{5}\gamma_{\mu}=-\gamma_{\mu}\gamma_{5}.$ As a
result, we arrive at the following equations,%
\begin{equation}
\frac{1}{2}e_{\mu\nu\sigma\tau}\partial^{\nu}\left(
\gamma^{\sigma}\gamma^{\tau}\psi\right) =\left(
i\partial_{\mu}+m\gamma_{\mu}\right) \gamma_{5}\psi,  \label{PauliLubanski}
\end{equation}
with summation over any two repeated indices.

The latter equation can also be obtained in view of (\ref{Generators}), by
letting $p_{\mu }=i\partial _{\mu }$ in operator relation (\ref%
{PauliLubanskiDirac}). Once again, our goal is to emphasize that both
overdetermined systems (\ref{DiracEquationSigma}) and (\ref{PauliLubanski}),
which are related to the Pauli-Luba\'{n}ski vector, are equivalent to
Dirac's equation in vacuum (\ref{DiracEquationGamma}).

In components, by (\ref{PauliLubanski}), for the rotations $\boldsymbol{%
\Sigma }$ and boosts $\boldsymbol{\alpha}$ the following standard equations
hold%
\begin{equation}
i\left( \nabla\cdot\boldsymbol{\Sigma}\right) \psi=\left( i\partial
_{0}+m\gamma_{0}\right) \gamma_{5}\psi  \label{KLSzero}
\end{equation}
and%
\begin{equation}
i\partial_{0}\mathbf{\Sigma}\psi-\left( \nabla\times\boldsymbol{\alpha }%
\right) \psi=\left( i\nabla-m\boldsymbol{\gamma}\right) \gamma_{5}\psi,
\label{KLSvector}
\end{equation}
respectively. Once again, this system is overdetermined and by a proper
matrix multiplication, each of the four equations (\ref{KLSzero})--(\ref%
{KLSvector}) can be reduced to a single Dirac's equation. We leave the
details to the reader.

\subsection{Relativistic definition of spin for Dirac particles}

In view of (\ref{PauliLubanskiDirac}) and (\ref{MassSpin}), one gets%
\begin{align*}
4w_{\mu}w^{\mu} & =\left( p_{\mu}+m\gamma_{\mu}\right) \gamma_{5}\left(
p^{\mu}+m\gamma^{\mu}\right) \gamma_{5} \\
& =\left( p_{\mu}+m\gamma_{\mu}\right) \gamma_{5}^{2}\left( p^{\mu
}-m\gamma^{\mu}\right) \\
& =p_{\mu}p^{\mu}-m^{2}\gamma_{\mu}\gamma^{\mu}=-3m^{2},
\end{align*}
where $s\left( s+1\right) =3/4,$ or $s=1/2,$ in covariant form.

On the other hand, introducing the familiar generators $\boldsymbol{M}%
=\left( i/2\right) \boldsymbol{\Sigma }$ and $\boldsymbol{N}=\left(
1/2\right) \boldsymbol{\alpha }$ for the rotations and boosts, respectively,
one gets%
\begin{equation}
\boldsymbol{N}^{2}=-\boldsymbol{M}^{2}=3/4,\qquad \boldsymbol{M}\cdot
\boldsymbol{N}=i\left( 3/4\right) \gamma ^{5}.  \label{CasimirLorentz}
\end{equation}%
In view of $\boldsymbol{\Sigma }\gamma ^{5}=\gamma ^{5}\boldsymbol{\Sigma }=%
\boldsymbol{\alpha }$ $,$ in the complex space of the bispinors under
consideration, we arrive at%
\begin{equation}
\boldsymbol{M}\mathbf{\pm }i\boldsymbol{N=}\frac{i}{2}\left( 1\mp \gamma
_{5}\right) \boldsymbol{\Sigma }  \label{MplusminusN}
\end{equation}%
and the Casimir operators of the proper orthochronous Lorentz group are
given by $\left( \boldsymbol{M}\mathbf{\pm }i\boldsymbol{N}\right)
^{2}/4=-3\left( 1\mp \gamma _{5}\right) ^{2}/16=-3\left( 1\mp \gamma
_{5}\right) /8.$ Here, $\boldsymbol{M}^{2}=-s\left( s+1\right) =-3/4,$ which
implies, once again, that the spin is equal to $1/2$ (we have chosen
real-valued boost generators; see also \cite{Bargmann47}, \cite%
{BargmannWigner48}, \cite{BarutRaczka80}, \cite{Bogolubovetal90}, \cite%
{Gelfandetal}, \cite{RumerFetQFT}, \cite{Schweber61} and \cite{Wein} for
more details on the Lorentz group representations).

\textit{Miscellaneous. }In addition,%
\begin{equation}
\frac{1}{2}e_{\mu \nu \sigma \tau }\partial ^{\nu }\left( \gamma ^{\mu
}\gamma ^{\sigma }\gamma ^{\tau }\psi \right) =3i\gamma ^{5}\gamma _{\nu
}\partial ^{\nu }\psi =3m\gamma ^{5}\psi ,  \label{Mis1}
\end{equation}%
in view of familiar relations:%
\begin{equation}
\gamma ^{5}=\frac{i}{4!}e_{\mu \nu \sigma \tau }\gamma ^{\mu }\gamma ^{\nu
}\gamma ^{\sigma }\gamma ^{\tau },\qquad \gamma ^{5}\gamma _{\mu }=\frac{i}{%
3!}e_{\mu \nu \sigma \tau }\gamma ^{\nu }\gamma ^{\sigma }\gamma ^{\tau }
\label{Mis2}
\end{equation}%
(see, for example, \cite{MoskalevQFT}).

\subsection{Massless limit}

Under the transformation,%
\begin{eqnarray}
\gamma ^{\mu }\rightarrow \gamma ^{\prime \mu } &=&U\gamma ^{\mu
}U^{-1},\qquad \psi \rightarrow \psi ^{\prime }=\left(
\begin{array}{c}
\phi \\
\chi%
\end{array}%
\right) =U\psi ,  \label{WeylGamma} \\
\qquad \qquad U &=&\frac{1}{\sqrt{2}}\left(
\begin{array}{cc}
I & I \\
I & -I%
\end{array}%
\right) =U^{-1},  \notag
\end{eqnarray}%
Dirac's equation (\ref{DiracEquationGamma}) takes a familiar block form,%
\begin{equation}
i\left(
\begin{array}{cc}
0 & \partial _{0}-\boldsymbol{\sigma }\cdot \nabla \\
\partial _{0}+\boldsymbol{\sigma }\cdot \nabla & 0%
\end{array}%
\right) \left(
\begin{array}{c}
\phi \\
\chi%
\end{array}%
\right) =m\left(
\begin{array}{c}
\phi \\
\chi%
\end{array}%
\right)  \label{DiracWeyl}
\end{equation}
(see, for example, \cite{Feynman}, \cite{MoskalevQFT}, and \cite%
{PeskinSchroeder}). As $m\rightarrow 0,$ this system decouples,%
\begin{equation}
\partial _{0}\phi +\left( \boldsymbol{\sigma }\cdot \nabla \right) \phi
=0,\qquad \partial _{0}\chi -\left( \boldsymbol{\sigma }\cdot \nabla \right)
\chi =0,  \label{WeylDiracLimit}
\end{equation}%
resulting in Weyl's two-component equations for massless neutrinos.

\section{Weyl equation for massless neutrinos}

The complex unimodular matrix group $SL\left( 2,%
\mathbb{C}
\right) $ is a two-fold universal covering group of the Lorentz group $%
SO\left( 1,3\right) $ (see, for example, \cite{BarutRaczka80}, \cite%
{Bogolubovetal90}, \cite{Gelfandetal}, \cite{RumerFetQFT}). In this section,
we shall use this connection in order to analyze the two-component spinor
field associated with Weyl's equation.

\subsection{Rotations, boosts, and their generators}

Let us consider the fundamental representation of $SL\left( 2,%
\mathbb{C}
\right) ,$ namely, we take a spinor field,
\begin{equation}
\phi \left( x\right) =\left(
\begin{array}{c}
\phi _{1} \\
\phi _{2}%
\end{array}%
\right) \in \left.
\mathbb{C}
\right. ^{2},  \label{2spinor}
\end{equation}%
and postulate the transformation law under the proper orthochronous Lorentz
group as follows%
\begin{equation}
\phi ^{\prime }\left( x^{\prime }\right) =S_{\Lambda }\phi \left( x\right)
,\qquad x^{\prime }=\Lambda x,\qquad S_{\Lambda }=\exp \left( \frac{1}{4}%
\theta _{\mu \nu }\Sigma ^{\mu \nu }\right) .  \label{SpinorTransform2}
\end{equation}%
Explicitly, these transformations include rotations,%
\begin{equation}
S_{R}=e^{i\theta \left( \boldsymbol{n\cdot \sigma }\right) /2}=\cos \frac{%
\theta }{2}+i\left( \boldsymbol{n\cdot \sigma }\right) \sin \frac{\theta }{2}%
,\qquad \left( \boldsymbol{n\cdot \sigma }\right) ^{2}=I
\label{RotationsSpinor}
\end{equation}%
about the coordinate axes $\boldsymbol{n}=\left\{ \mathbf{e}_{1},\mathbf{e}%
_{2},\mathbf{e}_{3}\right\} ,$ and boosts,%
\begin{equation}
S_{L}=e^{-\vartheta \left( \boldsymbol{n\cdot \sigma }\right) /2}=\cosh
\frac{\vartheta }{2}-\left( \boldsymbol{n\cdot \sigma }\right) \sinh \frac{%
\vartheta }{2},\qquad \boldsymbol{n}=\frac{\boldsymbol{v}}{v}
\label{BoostsSpinor}
\end{equation}%
in the directions $\boldsymbol{n}=\left\{ \mathbf{e}_{1},\mathbf{e}_{2},%
\mathbf{e}_{3}\right\} ,$ respectively, when the familiar relations,%
\begin{equation}
v=\tanh \vartheta ,\qquad \cosh \vartheta =\frac{1}{\sqrt{1-v^{2}}},\qquad
\sinh \vartheta =\frac{v}{\sqrt{1-v^{2}}}\quad \left( c=1\right) ,
\label{BoostParameters}
\end{equation}%
hold. (Here, $\sigma _{1},$ $\sigma _{2},$ $\sigma _{3}$ are the Pauli
matrices with the products given by $\sigma _{p}\sigma _{q}=ie_{pqr}\sigma
_{r}+\delta _{pq}$ and $\left\{ \mathbf{e}_{a}\right\} _{a=1,2,3}$ is an
orthonormal basis in $\left.
\mathbb{R}
\right. ^{3}.)$

The action of the generators $M^{\alpha\beta}=x^{\beta}\partial^{\alpha
}-x^{\alpha}\partial^{\beta}$ takes the form%
\begin{align}
M^{\alpha\beta}\phi\left( x\right) & :=-\left. \left( \frac{d}{%
d\theta_{\alpha\beta}}\phi^{\prime}\left( \Lambda x\right) \right)
\right\vert _{\theta_{\alpha\beta}=0}  \label{2SpinorGenerators} \\
& =-\left. \left( \frac{dS}{d\theta_{\alpha\beta}}\right) \right\vert
_{\theta_{\alpha\beta}=0}\phi\left( x\right) =-\frac{1}{2}\Sigma
^{\alpha\beta}\phi\left( x\right)  \notag
\end{align}
for the corresponding one-parameter subgroups: $S_{\Lambda}=\exp\left(
\theta_{\alpha\beta}\Sigma^{\alpha\beta}/2\right) $ $(\alpha,\beta=0,1,2,3$
are fixed with no summation).

In block form, the generators of this spinor representation are given by%
\footnote{%
Another choice of the generators in the transformation laws (\ref%
{SpinorTransform2})--(\ref{BoostsSpinor}), corresponds to $\vartheta
\rightarrow-\vartheta;$ see, for example, \cite{Schweber61}.}%
\begin{equation}
\Sigma^{\alpha\beta}=-\Sigma^{\beta\alpha}=\left(
\begin{array}{cc}
0 & -\sigma_{q}\medskip \\
\sigma_{p} & ie_{pqr}\sigma_{r}%
\end{array}
\right) =\left(
\begin{array}{cccc}
0 & -\sigma_{1} & -\sigma_{2} & -\sigma_{3} \\
\sigma_{1} & 0 & i\sigma_{3} & -i\sigma_{2} \\
\sigma_{2} & -i\sigma_{3} & 0 & i\sigma_{1} \\
\sigma_{3} & i\sigma_{2} & -i\sigma_{1} & 0%
\end{array}
\right)  \label{2generators}
\end{equation}
and the following self-duality identity holds%
\begin{equation}
e_{\mu\nu\sigma\tau}\Sigma^{\sigma\tau}=2i\Sigma_{\mu\nu}=2ig_{\mu\sigma
}g_{\nu\tau}\Sigma^{\sigma\tau},  \label{2SpinorSelfDuality}
\end{equation}
where%
\begin{equation}
\Sigma_{\alpha\beta}=g_{\alpha\mu}g_{\beta\nu}\Sigma^{\mu\nu}=\left(
\begin{array}{cc}
0 & \sigma_{q}\medskip \\
-\sigma_{p} & ie_{pqr}\sigma_{r}%
\end{array}
\right) =\left(
\begin{array}{cccc}
0 & \sigma_{1} & \sigma_{2} & \sigma_{3} \\
-\sigma_{1} & 0 & i\sigma_{3} & -i\sigma_{2} \\
-\sigma_{2} & -i\sigma_{3} & 0 & i\sigma_{1} \\
-\sigma_{3} & i\sigma_{2} & -i\sigma_{1} & 0%
\end{array}
\right)  \label{2generatorsdown}
\end{equation}
stated here for the reader's convenience.

\subsection{The Pauli-Luba\'{n}ski vector and Weyl's equation}

Letting $\lambda =-1/2$ in (\ref{WrongHelicity}), one gets%
\begin{equation}
e_{\mu \nu \sigma \tau }\left( i\partial ^{\nu }\right) \left( -iM^{\sigma
\tau }\right) \phi =-\frac{1}{2}e_{\mu \nu \sigma \tau }\partial ^{\nu
}\Sigma ^{\sigma \tau }\phi =-i\partial _{\mu }\phi
\label{PauliLubanskiWeyl}
\end{equation}%
by (\ref{2SpinorGenerators}). With the help of (\ref{2SpinorSelfDuality}),
we finally arrive at the following overdetermined system:%
\begin{equation}
\Sigma ^{\mu \nu }\partial _{\nu }\phi =\partial ^{\mu }\phi ,
\label{WeylSigma}
\end{equation}%
which takes the explicit form,%
\begin{align}
-\left( \boldsymbol{\sigma }\cdot \nabla \right) \phi & =\partial _{0}\phi ,
\label{WeylEquation} \\
\partial _{0}\boldsymbol{\sigma }\phi +i\left( \nabla \times \boldsymbol{%
\sigma }\right) \phi & =-\nabla \phi .  \notag
\end{align}%
The latter vectorial equation can be obtained from the first one by matrix
multiplication with the help of familiar products of the Pauli matrices.
Moreover, in (\ref{WrongHelicity}), only the value $\lambda =-1/2$ results
in a consistent system, defining the spin as $s=\left\vert \lambda
\right\vert =1/2.$

Thus, the relativistic two-component Weyl equations for a massless particle
with the spin $1/2,$ namely,$\ \partial _{0}\phi +\left( \boldsymbol{\sigma }%
\cdot \nabla \right) \phi =0,$ can be derived from the representation theory
of the Poincar\'{e} group with the aid of the Pauli-Luba\'{n}ski vector%
\footnote{%
Some authors, see for example, \cite{Ryder}, \cite{ScheckQuantumPhysics},
suggest that an empirical condition is required in order to quantize the
value of spin for a massless particle by (\ref{WrongHelicity}). As we have
just demonstrated, for Weyl's equation, it is a misconception. Indeed, the
full relativistic analysis automatically includes the quantization rule of
the corresponding spin and helicity as a consistency condition of the
overdetermined system (\ref{WeylSigma}).}.

\subsection{Covariance}

Equations (\ref{WeylSigma}) are covariant under a proper Lorentz
transformation. In view of the laws (\ref{SpinorTransform2}) one gets%
\begin{equation}
S_{\Lambda}^{-1}\Sigma^{\sigma\tau}S_{\Lambda}=\Lambda_{\
\mu}^{\sigma}\Lambda_{\ \nu}^{\tau}\Sigma^{\mu\nu}  \label{TensorTransform}
\end{equation}
and%
\begin{equation}
\frac{1}{2}\left[ \Sigma^{\alpha\beta},\ \Sigma^{\gamma\delta}\right]
=g^{\alpha\gamma}\Sigma^{\beta\delta}-g^{\alpha\delta}\Sigma^{\beta\gamma
}+g^{\beta\delta}\Sigma^{\alpha\gamma}-g^{\beta\gamma}\Sigma^{\alpha\delta}.
\label{CommutatorWeyl}
\end{equation}
(cf. Sections 5.2--5.3).

On the second hand, with the help of the Pauli-Luba\'{n}ski vector, from (%
\ref{PauliLubanskiWeyl}) one can get,%
\begin{equation}
\frac{1}{2}e_{\mu \nu \sigma \tau }\partial ^{\prime \nu }\Sigma ^{\sigma
\tau }\phi ^{\prime }=i\partial _{\mu }^{\prime }\phi ^{\prime },
\label{PauliLubanskiWeylNew}
\end{equation}%
in the new system of coordinates, when $\phi ^{\prime }\left( x^{\prime
}\right) =S\phi \left( x\right) $ and $x^{\prime }=\Lambda x.$ Let us
multiply both sides of the latter equation by $S^{-1}\Lambda _{\ \lambda
}^{\mu }$ from the left and then use (\ref{TensorTransform}) together with
the two convenient transformations,%
\begin{equation}
S\partial _{\lambda }\phi =\Lambda _{\ \lambda }^{\mu }\partial _{\mu
}^{\prime }\phi ^{\prime },\qquad \partial ^{\prime \nu }\phi ^{\prime
}=\Lambda _{\ \kappa }^{\nu }S\partial ^{\kappa }\phi ,
\label{WeylIdentities}
\end{equation}%
which can be readily verified with the aid of%
\begin{equation}
\left( \Lambda ^{-1}\right) _{\ \nu }^{\lambda }\Lambda _{\ \tau }^{\nu
}=\delta _{\tau }^{\lambda },\qquad \left( \Lambda ^{-1}\right) _{\ \nu
}^{\lambda }=g^{\lambda \sigma }\Lambda _{\ \sigma }^{\mu }g_{\mu \nu }.
\label{LorentzOrthogonality}
\end{equation}%
As a result,%
\begin{equation}
\left( \det \Lambda \right) e_{\lambda \kappa \rho \chi }\partial ^{\kappa
}\Sigma ^{\rho \chi }\phi =i\partial _{\lambda }\phi ,
\label{PauliLubanskiWeylOld}
\end{equation}%
in view of the following determinant identity \cite{MoskalevQFT}:%
\begin{equation}
e_{\mu \nu \sigma \tau }\Lambda _{\ \lambda }^{\mu }\Lambda _{\ \kappa
}^{\nu }\Lambda _{\ \rho }^{\sigma }\Lambda _{\ \chi }^{\tau }=\left( \det
\Lambda \right) e_{\lambda \kappa \rho \chi }.
\end{equation}%
This consideration reveals the pseudo-vector character of the Pauli-Luba\'{n}%
ski operator.

\textit{Note.} Weyl's equation (\ref{WeylEquation}) was originally
introduced in \cite{Weyl} and then quickly rejected \cite{Pauli41}, being
\textquotedblleft resurrected\textquotedblright\ only after the discovery of
parity violation in beta decay \cite{Roy01}, \cite{Wuetal57}. (The
experimentally detected oscillation among the different flavors of neutrinos
leads us to believe that they are not massless after all \cite{Kajita15},
\cite{McDonald15}, \cite{Pontecorvo83}.)

Let $\boldsymbol{v},$ the velocity vector of the moving frame of reference,
lie along one of the coordinate axes, say $\left\{ \mathbf{e}_{a}\right\}
_{a=1,2,3}.$ Also, let $\boldsymbol{\sigma }=\boldsymbol{\sigma }_{\parallel
}+\boldsymbol{\sigma }_{\perp }$ be the corresponding \textquotedblleft
orthogonal decomposition\textquotedblright\ in parallel and perpendicular
directions, respectively. These components transform as,
\begin{equation}
S_{\Lambda }^{-1}\boldsymbol{\sigma }_{\parallel }S_{\Lambda }=\boldsymbol{%
\sigma }_{\parallel },\qquad S_{\Lambda }^{-1}\boldsymbol{\sigma }_{\perp
}S_{\Lambda }=\frac{\boldsymbol{\sigma }_{\perp }-i\left( \boldsymbol{v}%
\times \boldsymbol{\sigma }\right) }{\sqrt{1-v^{2}}},  \label{SigmaTransform}
\end{equation}%
under a Lorentz transformation, thus resembling the transformations of
electromagnetic fields\ in classical electrodynamics \cite{KrLanSus150},
\cite{KrLanSu15}, \cite{MinkowskiI}, \cite{SchweberBetheHoffmann}, \cite%
{ToptyginI}.

\noindent \textbf{Example.} Let $\boldsymbol{v}=v\mathbf{e}_{1}.$ Then $%
\boldsymbol{\sigma }_{\parallel }=\sigma _{1},$ $\boldsymbol{\sigma }_{\perp
}=\left\{ \sigma _{2},\sigma _{3}\right\} $ and%
\begin{equation}
S_{L}=e^{-\left( \vartheta /2\right) \sigma _{1}}=\cosh \frac{\vartheta }{2}%
-\sigma _{1}\sinh \frac{\vartheta }{2},\qquad S_{L}^{-1}=e^{\left( \vartheta
/2\right) \sigma _{1}}=\cosh \frac{\vartheta }{2}+\sigma _{1}\sinh \frac{%
\vartheta }{2}.
\end{equation}%
By the transformation law (\ref{SigmaTransform}), one should get $%
S^{-1}\sigma _{1}S=\sigma _{1},$ which is obvious, and%
\begin{equation}
S^{-1}\sigma _{2}S=\frac{\sigma _{2}+iv\sigma _{3}}{\sqrt{1-v^{2}}},\qquad
S^{-1}\sigma _{3}S=\frac{\sigma _{3}-iv\sigma _{2}}{\sqrt{1-v^{2}}}
\end{equation}%
for the corresponding Lorentz boost. Let us directly verify, for instance,
the first relation. Indeed,%
\begin{align*}
S^{-1}\sigma _{2}S& =\left( \cosh \frac{\vartheta }{2}+\sigma _{1}\sinh
\frac{\vartheta }{2}\right) \sigma _{2}\left( \cosh \frac{\vartheta }{2}%
-\sigma _{1}\sinh \frac{\vartheta }{2}\right) \\
& =\sigma _{2}\cosh ^{2}\frac{\vartheta }{2}+\left( \sigma _{1}\sigma
_{2}-\sigma _{2}\sigma _{1}\right) \cosh \frac{\vartheta }{2}\sinh \frac{%
\vartheta }{2}-\sigma _{1}\sigma _{2}\sigma _{1}\sinh ^{2}\frac{\vartheta }{2%
} \\
& =\sigma _{2}\cosh \vartheta +i\sigma _{3}\sinh \vartheta =\frac{\sigma
_{2}+iv\sigma _{3}}{\sqrt{1-v^{2}}},
\end{align*}%
provided that $\tanh \vartheta =v.$ The proof of the last identity is
similar.

As is well-known, under spatial rotations the set of three Pauli matrices $%
\boldsymbol{\sigma }$ transform as a $3D$ vector. For instance,%
\begin{equation}
e^{-i\left( \theta /2\right) \sigma _{3}}\left(
\begin{array}{c}
\sigma _{1} \\
\sigma _{2} \\
\sigma _{3}%
\end{array}%
\right) e^{i\left( \theta /2\right) \sigma _{3}}=\left(
\begin{array}{ccc}
\cos \theta & \sin \theta & 0 \\
-\sin \theta & \cos \theta & 0 \\
0 & 0 & 1%
\end{array}%
\right) \left(
\begin{array}{c}
\sigma _{1} \\
\sigma _{2} \\
\sigma _{3}%
\end{array}%
\right) .
\end{equation}

\subsection{An alternative derivation}

Denoting, $\sigma ^{\mu }=\left( \sigma _{0}=I,\sigma _{1},\sigma
_{2},\sigma _{3}\right) ,$ one can rewrite Weyl's equation in a more
familiar form \cite{PeskinSchroeder}, \cite{Schweber61}:%
\begin{equation}
\sigma ^{\mu }\partial _{\mu }\phi =0.  \label{Weyl2}
\end{equation}%
Then, under the Lorentz transformations,%
\begin{equation}
S_{\Lambda }^{\dag }\sigma ^{\mu }S_{\Lambda }=\Lambda _{\ \nu }^{\mu
}\sigma ^{\nu }  \label{SigmaAlternative}
\end{equation}%
and%
\begin{equation}
\left( \Sigma ^{\alpha \beta }\right) ^{\dag }\sigma ^{\mu }+\sigma ^{\mu
}\Sigma ^{\alpha \beta }=2\left( g^{\beta \mu }\sigma ^{\alpha }-g^{\alpha
\mu }\sigma ^{\beta }\right) ,
\end{equation}%
which also implies the covariance\footnote{%
Relations of this spinor representation with Maxwell's equations are
discussed in section~6.2.}.

\noindent \textbf{Examples.} In particular, one can easily verify that%
\begin{equation}
e^{-\left( \vartheta /2\right) \sigma _{1}}\left(
\begin{array}{c}
\sigma _{0} \\
\sigma _{1} \\
\sigma _{2} \\
\sigma _{3}%
\end{array}%
\right) e^{-\left( \vartheta /2\right) \sigma _{1}}=\left(
\begin{array}{cccc}
\cosh \vartheta & -\sinh \vartheta & 0 & 0 \\
-\sinh \vartheta & \cosh \vartheta & 0 & 0 \\
0 & 0 & 1 & 0 \\
0 & 0 & 0 & 1%
\end{array}%
\right) \left(
\begin{array}{c}
\sigma _{0} \\
\sigma _{1} \\
\sigma _{2} \\
\sigma _{3}%
\end{array}%
\right) ,
\end{equation}%
with $\tanh \vartheta =v,$ and%
\begin{equation}
e^{-i\left( \theta /2\right) \sigma _{3}}\left(
\begin{array}{c}
\sigma _{0} \\
\sigma _{1} \\
\sigma _{2} \\
\sigma _{3}%
\end{array}%
\right) e^{i\left( \theta /2\right) \sigma _{3}}=\left(
\begin{array}{cccc}
1 & 0 & 0 & 0 \\
0 & \cos \theta & \sin \theta & 0 \\
0 & -\sin \theta & \cos \theta & 0 \\
0 & 0 & 0 & 1%
\end{array}%
\right) \left(
\begin{array}{c}
\sigma _{0} \\
\sigma _{1} \\
\sigma _{2} \\
\sigma _{3}%
\end{array}%
\right)
\end{equation}%
for the corresponding boost and rotation, respectively.

\section{Proca equation}

The fundamental four-vector representation of the proper orthochronous
Lorentz group $SO_{+}\left( 1,3\right) $ is related to the relativistic wave
equation for a massive particle with spin $1.$

\subsection{Massive vector field}

The relativistic equation of motion for a real or complex four-vector field $%
A^{\mu }=\left( A^{0},\mathbf{A}\right) $ with a positive mass $m>0$ can be
derived in a natural way with the help of the Pauli-Luba\'{n}ski vector. By
definition,%
\begin{align}
w_{\mu }A^{\alpha }& =\frac{1}{2}e_{\mu \nu \sigma \tau }\partial ^{\nu
}M^{\sigma \tau }A^{\alpha }=\frac{1}{2}e_{\mu \nu \sigma \tau }\partial
^{\nu }\left( \left( m^{\sigma \tau }\right) _{\rho }^{\alpha }A^{\rho
}\right) \\
& =\frac{1}{2}e_{\mu \nu \sigma \tau }\partial ^{\nu }\left( g^{\sigma
\alpha }A^{\tau }-g^{\tau \alpha }A^{\sigma }\right) =-g^{\alpha \nu }e_{\mu
\nu \sigma \tau }\partial ^{\sigma }A^{\tau },  \notag
\end{align}%
where the matrix form of the generators (\ref{OneParameter}) has been used.
The standard decomposition,%
\begin{equation*}
\partial ^{\mu }A^{\nu }=\frac{1}{2}\left( \partial ^{\mu }A^{\nu }-\partial
^{\nu }A^{\mu }\right) +\frac{1}{2}\left( \partial ^{\mu }A^{\nu }+\partial
^{\nu }A^{\mu }\right) ,
\end{equation*}%
followed by the dual tensor relation,%
\begin{equation}
F^{\mu \nu }=\partial ^{\mu }A^{\nu }-\partial ^{\nu }A^{\mu },\qquad e_{\mu
\nu \sigma \tau }F^{\sigma \tau }=-2G_{\mu \nu },  \label{FTensorProca}
\end{equation}%
gives the explicit action of the Pauli-Luba\'{n}ski operator on the
four-vector field:%
\begin{equation}
w_{\mu }A^{\alpha }=g^{\alpha \nu }G_{\mu \nu }=g_{\mu \nu }G^{\nu \alpha
},\qquad w_{\mu }A_{\alpha }=G_{\mu \alpha }.  \label{PLProcaOne}
\end{equation}%
It is worth noting that this results in a second rank four-tensor.

In a similar fashion, for the squared operator,%
\begin{equation}
w^{2}A^{\alpha}=w^{\mu}\left( w_{\mu}A^{\alpha}\right) =\frac{1}{2}e_{\mu
\nu\sigma\tau}\partial^{\nu}\left( M^{\sigma\tau}G^{\mu\alpha}\right) .
\end{equation}
But, in view of (4.7) of \cite{KrLanSus150},%
\begin{equation}
M^{\sigma\tau}G^{\mu\alpha}=g^{\sigma\mu}G^{\tau\alpha}-g^{\tau\mu}G^{\sigma%
\alpha}+g^{\sigma\alpha}G^{\mu\tau}-g^{\tau\alpha}G^{\mu\sigma},
\label{MTensorG}
\end{equation}
and with the help of a companion dual tensor identity, $e^{\mu\nu\sigma\tau
}G_{\sigma\tau}=2F^{\mu\nu},$ one gets
\begin{equation*}
w^{2}A^{\alpha}=g^{\sigma\alpha}\partial^{\nu}\left( e_{\nu\sigma\mu\tau
}G^{\mu\tau}\right) =2g^{\sigma\alpha}\partial^{\nu}\left( F_{\nu\sigma
}\right) =2\partial_{\nu}F^{\nu\alpha}=-2m^{2}A^{\alpha}
\end{equation*}
as \ a consequence of (\ref{MassSpin}). As a result, we have arrived at the
Proca equation for a vector particle with a finite mass \cite{Proca36},%
\begin{equation}
\partial_{\nu}F^{\nu\mu}+m^{2}A^{\mu}=0,\qquad
F^{\mu\nu}=\partial^{\mu}A^{\nu}-\partial^{\nu}A^{\mu},
\label{ProcaEquation}
\end{equation}
directly from the representation theory of the Poincar\'{e} group. In view
of $w^{2}=-m^{2}s\left( s+1\right) =-2m^{2},$ one concludes that the spin of
the particle is equal to one.

Moreover,%
\begin{equation}
\partial_{\mu}A^{\mu}=0,\qquad\square A^{\mu}=-m^{2}A^{\mu},\quad m>0,
\end{equation}
in view of%
\begin{equation}
0\equiv\partial_{\mu}\partial_{\nu}F^{\nu\mu}=-m^{2}\partial_{\mu}A^{\mu}
\end{equation}
and%
\begin{equation}
\partial_{\nu}F^{\nu\mu}=\partial_{\nu}\left(
\partial^{\nu}A^{\mu}-\partial^{\mu}A^{\nu}\right)
=\partial_{\nu}\partial^{\nu}A^{\mu}=-m^{2}A^{\mu}.
\end{equation}
The massless case of the Proca equation, $m=0,$ reveals a gauge invariance.
If $A_{\mu}\rightarrow A_{\mu}^{\prime}=A_{\mu}+\partial_{\mu}f,$ then $%
F_{\mu \nu}^{\prime}=\partial_{\mu}\left( A_{\nu}+\partial_{\nu}f\right)
-\partial_{\nu}\left( A_{\mu}+\partial_{\mu}f\right) =F_{\mu\nu}.$

\subsection{An alternative \textquotedblleft bispinor\textquotedblright\
derivation}

Let us consider a second rank bispinor of the form\footnote{%
This is the so-called \textquotedblleft Feynman slash\textquotedblright\
notation from quantum electrodynamics \cite{Feynman}.} $Q=A_{\lambda }\gamma
^{\lambda }=A^{\lambda }\gamma _{\lambda },$ where $\left\{ \gamma ^{\lambda
}\right\} _{\lambda =0,1,2,3}$ are the standard gamma matrices. We shall use
the following transformation law for a proper Lorentz transformation,
\begin{equation}
Q^{\prime }\left( x^{\prime }\right) =S_{\Lambda }Q\left( x\right)
S_{\Lambda }^{-1},\qquad x^{\prime }=\Lambda x,  \label{ProcaBispinor2}
\end{equation}%
where the matrix $S_{\Lambda }$ is given by (\ref{MatrixS}), as in the case
of Dirac's equation. Then $A^{\lambda }$ must be a four-vector. Indeed,
\begin{align}
Q^{\prime }=A_{\lambda }\delta _{\sigma }^{\lambda }\left( S\gamma ^{\sigma
}S^{-1}\right) & =\left( g^{\tau \lambda }A_{\lambda }\Lambda _{\ \tau
}^{\nu }\right) g_{\mu \nu }\left[ \Lambda _{\ \sigma }^{\mu }\left( S\gamma
^{\sigma }S^{-1}\right) \right] =A^{\prime \nu }\gamma _{\nu },
\label{ProcaBispinor2Tr} \\
\qquad \qquad \qquad A^{\prime \nu }& =\Lambda _{\ \tau }^{\nu }A^{\tau },
\notag
\end{align}%
in view of an \textquotedblleft inversion\textquotedblright\ of (\ref%
{TransformationLawGamma}), $\Lambda _{\ \sigma }^{\mu }\left( S\gamma
^{\sigma }S^{-1}\right) =\gamma ^{\mu },$ and the familiar property: $A_{\mu
}B^{\mu }=\mathrm{invariant}$ or%
\begin{equation}
g_{\mu \nu }\Lambda _{\ \sigma }^{\mu }\Lambda _{\ \tau }^{\nu }g^{\tau
\lambda }=\delta _{\sigma }^{\lambda }.  \label{LorentzOrthogonalityTwo}
\end{equation}

In this \textquotedblleft bispinor representation\textquotedblright, the
action of generators of the corresponding one-parameter subgroups is given by%
\begin{equation}
M^{\alpha\beta}Q:=-\left. \left( \frac{d}{d\theta_{\alpha\beta}}Q^{\prime
}\left( \Lambda x\right) \right) \right\vert _{\theta_{\alpha\beta}=0}=\frac{%
1}{2}\left( \Sigma^{\alpha\beta}Q-Q\Sigma^{\alpha\beta}\right) .
\end{equation}
By letting $Q=\gamma^{\lambda}A_{\lambda},$ one gets:%
\begin{equation*}
M^{\alpha\beta}Q=\left( g^{\beta\lambda}\gamma^{\alpha}-g^{\alpha\lambda
}\gamma^{\beta}\right) A_{\lambda},
\end{equation*}
in view of the familiar commutator (\ref{CommutatorSigmaGamma}). As a
result, we obtain $w_{\mu}Q=G_{\mu\tau}\gamma^{\tau}$ for the action of the
Pauli-Luba\'{n}ski operator on the bispinor $Q,$ which also follows from (%
\ref{PLProcaOne}). One gets, in a similar fashion, that%
\begin{equation}
w^{2}Q=\frac{1}{2}e_{\mu\nu\sigma\tau}\partial^{\nu}\left( M^{\sigma\tau
}G^{\mu\lambda}\right) \gamma_{\lambda},
\end{equation}
where equation (\ref{MTensorG}) holds, once again, in view of the
transformation law of the four-tensor $G^{\mu\lambda}.$ As a result, Proca's
equation follows.

\subsection{Maxwell's equations vs Proca equation}

For the real-valued vector potential $A^{\mu }$ and $m=0,$ the Proca
equation (\ref{ProcaEquation}) is reduced to the Maxwell equations in
vacuum. Indeed, in view of the dual relation,%
\begin{equation}
6\partial _{\nu }G^{\mu \nu }=-e^{\mu \nu \sigma \tau }\left( \partial _{\nu
}F_{\sigma \tau }+\partial _{\sigma }F_{\tau \nu }+\partial _{\tau }F_{\nu
\sigma }\right) =0,
\end{equation}%
both pairs of Maxwell's equations can be written together in the following
complex form%
\begin{equation}
\partial _{\nu }Q^{\mu \nu }=0,\qquad Q^{\mu \nu }=F^{\mu \nu }-\frac{i}{2}%
e^{\mu \nu \sigma \tau }F_{\sigma \tau },  \label{4TensorQ}
\end{equation}%
with the help of a self-dual complex four-tensor \cite{Bia:Bia75}, \cite%
{KrLanSus150}:%
\begin{equation}
2iQ^{\mu \nu }=e^{\mu \nu \sigma \tau }Q_{\sigma \tau },\quad e_{\mu \nu
\sigma \tau }Q^{\sigma \tau }=2iQ_{\mu \nu }.  \label{SelfDualQ}
\end{equation}%
The corresponding overdetermined system of Maxwell's equations in vacuum,
which is related to the Pauli-Luba\'{n}ski vector, is investigated in \cite%
{KrLanSus150}; see equation (5.2) there and section 5.5 below. (Two
different spinor forms of Maxwell's equation will be discussed in section~6.)

\section{Complex vector field}

Finally, we would like to discuss the fundamental representation for the
complex orthogonal group $SO\left( 3,%
\mathbb{C}
\right) $ in connection with Maxwell's equations in vacuum.

\subsection{Vector covariant form}

As is well-known, the transformation laws of the complex electromagnetic
field $\mathbf{F}\left( \boldsymbol{r},t\right) =\mathbf{E}+i\mathbf{H}\in
\left.
\mathbb{C}
\right. ^{3}$ in vacuum can be written in terms of $SO\left( 3,%
\mathbb{C}
\right) $ rotations. In addition to the standard rotations of the frame of
reference, the Lorentz transformations are equivalent to rotations through
imaginary angles thus preserving the relativistic invariant $\mathbf{F}^{2}=%
\mathbf{E}^{2}+\mathbf{H}^{2}+2i\mathbf{E}\cdot \mathbf{H},$ which can be
thought of as a \textquotedblleft complex length\textquotedblright\ of this
vector \cite{LanLif2}, \cite{MinkowskiI}. In these transformations, $\mathbf{%
F}^{\prime }\left( x^{\prime }\right) =S_{\Lambda }\mathbf{F}\left( x\right)
,$ $x^{\prime }=\Lambda x,$ (or $F_{p}^{\prime }\left( x^{\prime }\right)
=a_{pq}F_{q}\left( x\right) $ for a given complex orthogonal matrix), one
can choose $S_{R}=e^{-\omega \left( \mathbf{n}\cdot \mathbf{s}\right) }$ and
$S_{L}=e^{-i\upsilon \left( \mathbf{n}\cdot \mathbf{s}\right) }$ for the
rotations and boosts, respectively. Here, $\mathbf{n}=\left\{ \mathbf{e}_{1},%
\mathbf{e}_{2},\mathbf{e}_{3}\right\} ,$ when $\left\{ \mathbf{e}%
_{a}\right\} _{a=1,2,3}$ is an orthonormal basis in $\left.
\mathbb{R}
\right. ^{3},$ and $s_{1},$ $s_{2},$ $s_{3}$ are the real-valued spin
matrices \cite{VMK}:%
\begin{equation}
s_{1}=\left(
\begin{array}{ccc}
0 & 0 & 0 \\
0 & 0 & -1 \\
0 & 1 & 0%
\end{array}%
\right) ,\quad s_{2}=\left(
\begin{array}{ccc}
0 & 0 & 1 \\
0 & 0 & 0 \\
-1 & 0 & 0%
\end{array}%
\right) ,\quad s_{3}=\left(
\begin{array}{ccc}
0 & -1 & 0 \\
1 & 0 & 0 \\
0 & 0 & 0%
\end{array}%
\right) ,\quad  \label{SpinMatrices}
\end{equation}%
such that $\left[ s_{p},s_{q}\right] =s_{p}s_{q}-s_{q}s_{p}=e_{pqr}s_{r}$
and $s_{1}^{2}+s_{2}^{2}+s_{3}^{2}=-2.$ (In this representation the matrices
$\boldsymbol{M}=\mathbf{s}$ and $\boldsymbol{N}=i\mathbf{s}$ obey the
commutation law of the generators of the proper orthochronous Lorentz group.)

It can be directly verified that the corresponding generators,%
\begin{equation}
\Sigma^{\alpha\beta}=-\Sigma^{\beta\alpha}=\left(
\begin{array}{cc}
0 & is_{q}\medskip \\
-is_{p} & e_{pqr}s_{r}%
\end{array}
\right) =\left(
\begin{array}{cccc}
0 & is_{1} & is_{2} & is_{3} \\
-is_{1} & 0 & s_{3} & -s_{2} \\
-is_{2} & -s_{3} & 0 & s_{1} \\
-is_{3} & s_{2} & -s_{1} & 0%
\end{array}
\right) ,  \label{SigmaSpin}
\end{equation}
form a self-dual four-tensor%
\begin{equation}
2i\Sigma^{\mu\nu}=e^{\mu\nu\sigma\tau}\Sigma_{\sigma\tau},\quad e_{\mu
\nu\sigma\tau}\Sigma^{\sigma\tau}=2i\Sigma_{\mu\nu}.
\label{SelfDualitySO(3,C)}
\end{equation}
As a result, in this realization, $M^{\alpha\beta}\mathbf{F}=m^{\alpha\beta }%
\mathbf{F}$ and, in view of (\ref{WrongHelicity}) and (\ref%
{SelfDualitySO(3,C)}), we arrive at the set of overdetermined equations:%
\begin{equation}
\left( \Sigma^{\mu\nu}+g^{\mu\nu}\right) \partial_{\nu}\mathbf{F}=0,
\label{PLF4equation}
\end{equation}
where%
\begin{equation}
S_{\Lambda}^{-1}\Sigma^{\mu\nu}S_{\Lambda}=\Lambda_{\
\sigma}^{\mu}\Lambda_{\ \tau}^{\nu}\Sigma^{\sigma\tau},\qquad
x^{\prime\mu}=\Lambda _{\ \sigma}^{\mu}x^{\sigma},\qquad\mathbf{F}%
^{\prime}\left( x^{\prime }\right) =S_{\Lambda}\mathbf{F}\left( x\right)
\label{PLF4covariance}
\end{equation}
under a proper Lorentz transformation. Once again, the latter equations,
that, as we shall see later, determine the complete dynamics of the
electromagnetic field in vacuum, are obtained here by a pure
group-theoretical consideration up to an undetermined sign of the fixed
constant in (\ref{WrongHelicity}).

\subsection{Commutators}

For a one-parameter transformation in the plane $\left( \alpha,\beta\right)
, $ when%
\begin{equation}
\Lambda\left( \theta_{\alpha\beta}\right) =\exp\left( -\theta_{\alpha\beta
}\ m^{\alpha\beta}\right) ,\qquad\left( m^{\alpha\beta}\right) _{\nu}^{\mu
}=g^{\alpha\mu}\delta_{\nu}^{\beta}-g^{\beta\mu}\delta_{\nu}^{\alpha}
\end{equation}
and%
\begin{equation}
S_{\Lambda}=\exp\left( -\theta_{\alpha\beta}\Sigma^{\alpha\beta}\right)
,\qquad\theta_{\alpha\beta}=-\theta_{\beta\alpha}
\end{equation}
$(\alpha,\beta=0,1,2,3$ are fixed), one gets%
\begin{equation}
e^{\theta_{\alpha\beta}\Sigma^{\alpha\beta}}\Sigma^{\mu\nu}e^{-\theta
_{\alpha\beta}\Sigma^{\alpha\beta}}=\Lambda_{\ \sigma}^{\mu}\left(
\theta_{\alpha\beta}\right) \Lambda_{\ \tau}^{\nu}\left( \theta_{\alpha
\beta}\right) \Sigma^{\sigma\tau}.
\end{equation}
The differentiation $\left. \left( d/d\theta_{\alpha\beta}\right)
\right\vert _{\theta_{\alpha\beta}=0},$ results in a familiar law,%
\begin{align}
\left[ \Sigma^{\alpha\beta},\ \Sigma^{\mu\nu}\right] & :=\Sigma
^{\alpha\beta}\Sigma^{\mu\nu}-\Sigma^{\mu\nu}\Sigma^{\alpha\beta} \\
& =g^{\alpha\nu}\Sigma^{\beta\mu}-g^{\beta\nu}\Sigma^{\alpha\mu}+g^{\alpha
\mu}\Sigma^{\beta\nu}-g^{\beta\mu}\Sigma^{\alpha\nu},  \notag
\end{align}
which can be directly verified in components with the help of commutators of
the spin matrices (\ref{SpinMatrices}) that form the four-tensor (\ref%
{SigmaSpin}).

\subsection{Lorentz invariance}

With the help of (\ref{PLF4equation})--(\ref{PLF4covariance}), under a
Lorentz transformation,%
\begin{equation}
S^{-1}\left[ \left( \Sigma^{\prime\mu\nu}+g^{\prime\mu\nu}\right)
\partial_{\nu}^{\prime}\mathbf{F}^{\prime}\left( x^{\prime}\right) \right]
=0,
\end{equation}
where $\Sigma^{\prime\mu\nu}=\Sigma^{\mu\nu}=\mathrm{inv}$ and $g^{\prime
\mu\nu}=g^{\mu\nu}=\mathrm{inv}.$ Then%
\begin{align}
0 & =\left[ S^{-1}\left( \Sigma^{\mu\nu}+g^{\mu\nu}\right) S\right]
S^{-1}\partial_{\nu}^{\prime}\mathbf{F}^{\prime}\left( x^{\prime}\right) \\
& =\Lambda_{\ \sigma}^{\mu}\left( \Sigma^{\sigma\tau}+g^{\sigma\tau}\right)
S^{-1}\left[ \Lambda_{\ \tau}^{\nu}\partial_{\nu}^{\prime}\mathbf{F}^{\prime
}\left( x^{\prime}\right) \right]  \notag \\
& =\Lambda_{\ \sigma}^{\mu}\left( \Sigma^{\sigma\tau}+g^{\sigma\tau}\right)
S^{-1}S\partial_{\tau}\mathbf{F}\left( x\right)  \notag
\end{align}
by $\Lambda_{\ \tau}^{\nu}\partial_{\nu}^{\prime}\mathbf{F}^{\prime}\left(
x^{\prime}\right) =\partial_{\tau}\mathbf{F}\left( x\right) .$ Thus, our
equation $\left( \Sigma^{\mu\nu}+g^{\mu\nu}\right) \partial_{\nu}\mathbf{F}%
\left( x\right) =0$ preserves its covariant form for all real and complex
rotations $S\in SO\left( 3,%
\mathbb{C}
\right) .$

\subsection{Vector covariant form vs traditional form of Maxwell's equations}

A vector form of (\ref{PLF4equation}) is given by
\begin{equation}
\left( \nabla \cdot \mathbf{s}\right) \mathbf{F}=i\partial _{0}\mathbf{F}
\label{F0}
\end{equation}%
and%
\begin{equation}
\partial _{0}\mathbf{s}\ \mathbf{F}+i\left( \nabla \times \mathbf{s}\right)
\mathbf{F}=i\nabla \mathbf{F}.  \label{F123}
\end{equation}%
As it has been pointed out in \cite{KrLanSus150}, equation (\ref{F0})
implies the complex Maxwell equation, $\func{curl}\mathbf{F}=i\partial _{0}%
\mathbf{F}.$ The first component of (\ref{F123}) is given by $\partial
_{0}s_{1}\mathbf{F}+i\left( \partial _{2}s_{3}-\partial _{3}s_{2}\right)
\mathbf{F}=i\partial _{1}\mathbf{F},$ or%
\begin{align*}
& \partial _{0}\left(
\begin{array}{ccc}
0 & 0 & 0 \\
0 & 0 & -1 \\
0 & 1 & 0%
\end{array}%
\right) \left(
\begin{array}{c}
F_{1} \\
F_{2} \\
F_{3}%
\end{array}%
\right) +i\partial _{2}\left(
\begin{array}{ccc}
0 & -1 & 0 \\
1 & 0 & 0 \\
0 & 0 & 0%
\end{array}%
\right) \left(
\begin{array}{c}
F_{1} \\
F_{2} \\
F_{3}%
\end{array}%
\right) \\
& \ \ \ -i\partial _{3}\left(
\begin{array}{ccc}
0 & 0 & 1 \\
0 & 0 & 0 \\
-1 & 0 & 0%
\end{array}%
\right) \left(
\begin{array}{c}
F_{1} \\
F_{2} \\
F_{3}%
\end{array}%
\right) =i\partial _{1}\left(
\begin{array}{c}
F_{1} \\
F_{2} \\
F_{3}%
\end{array}%
\right)
\end{align*}%
and%
\begin{align*}
-i\partial _{2}F_{2}-i\partial _{3}F_{3}& =i\partial _{1}F_{1}, \\
-\partial _{0}F_{3}+i\partial _{2}F_{1}& =i\partial _{1}F_{2}, \\
\partial _{0}F_{2}+i\partial _{3}F_{1}& =i\partial _{1}F_{3}.
\end{align*}%
As a result, $\func{div}\mathbf{F}=0$ and $\left( \func{curl}\mathbf{F}%
\right) _{2,3}=i\partial _{0}\left( \mathbf{F}\right) _{2,3}.$ (Cyclic
permutations of the spatial indices cover the two remaining cases.)

\subsection{An alternative form of Maxwell's equations}

Complex covariant form of Maxwell's equations, which was introduced in \cite%
{LapUh31} (see also \cite{KrLanSus150}) , can be presented as follows%
\begin{equation}
\left( \partial _{0},\partial _{1},\partial _{2},\partial _{3}\right) \left(
\begin{array}{cccc}
0 & -F_{1} & -F_{2} & -F_{3} \\
F_{1} & 0 & iF_{3} & -iF_{2} \\
F_{2} & -iF_{3} & 0 & iF_{1} \\
F_{3} & iF_{2} & -iF_{1} & 0%
\end{array}%
\right) =\left( j_{0},j_{1},j_{2},j_{3}\right) ,  \label{MaxwellQuaterion}
\end{equation}%
or $\partial Q=j,$ by matrix multiplication. Here, $\mathcal{\partial }%
=\left( \partial _{0},\partial _{1},\partial _{2},\partial _{3}\right) ,$ $%
j=\left( j_{0},j_{1},j_{2},j_{3}\right) ,$ and $\mathbf{J}=\left(
J_{1},J_{2},J_{3}\right) $ such that%
\begin{eqnarray}
Q &=&\mathbf{F}\cdot \mathbf{J}=F_{1}\left(
\begin{array}{cccc}
0 & -1 & 0 & 0 \\
1 & 0 & 0 & 0 \\
0 & 0 & 0 & i \\
0 & 0 & -i & 0%
\end{array}%
\right)  \label{Quanternion} \\
&&+F_{2}\left(
\begin{array}{cccc}
0 & 0 & -1 & 0 \\
0 & 0 & 0 & -i \\
1 & 0 & 0 & 0 \\
0 & i & 0 & 0%
\end{array}%
\right) +F_{3}\left(
\begin{array}{cccc}
0 & 0 & 0 & -1 \\
0 & 0 & i & 0 \\
0 & -i & 0 & 0 \\
1 & 0 & 0 & 0%
\end{array}%
\right) .  \notag
\end{eqnarray}%
These matrices are transformed as a dual complex vector in $\left.
\mathbb{C}
\right. ^{3},$%
\begin{equation}
\Lambda J_{p}\Lambda ^{T}=a_{qp}J_{q},  \label{LorentzQuanternionTransform}
\end{equation}%
under a proper Lorentz transformation. In infinitesimal form,%
\begin{equation}
m^{\alpha \beta }\mathbf{J}+\mathbf{J}\left( m^{\alpha \beta }\right)
^{T}=\left( \Sigma ^{\alpha \beta }\right) ^{T}\mathbf{J},
\label{LorentzInfTransform}
\end{equation}%
which gives an alternative representation of the group $SO\left( 3,%
\mathbb{C}
\right) $ in a subspace of complex $4\times 4$ matrices. (Details are left
to the reader.)

As a result, $\Lambda Q\Lambda ^{T}=Q^{\prime }$ and, in \textquotedblleft
new\textquotedblright\ coordinates, $\partial ^{\prime }Q^{\prime
}=j^{\prime },$ provided that $\partial ^{\prime }=\partial \left( \Lambda
^{-1}\right) $ and $j^{\prime }=j\Lambda ^{T}.$

\section{On spinor forms of Maxwell's equations}

In conclusion, the complex matrix group $SL\left( 2,%
\mathbb{C}
\right) $ has a representation of the proper orthochronous Lorentz group $%
SO_{+}\left( 1,3\right) $ by the second rank spinors.

\subsection{Spinor covariant form}

The complex electromagnetic field in vacuum, $\mathbf{F}=\mathbf{E}+i\mathbf{%
H},$ can also be written in a familiar form of the following $2\times 2$
matrix:%
\begin{equation}
Q=\boldsymbol{\sigma }\cdot \mathbf{F}=\sigma _{1}F_{1}+\sigma
_{2}F_{2}+\sigma _{3}F_{3}=\left(
\begin{array}{cc}
F_{3} & F_{1}-iF_{2}\medskip \\
F_{1}+iF_{2}\medskip & -F_{3}%
\end{array}%
\right) ,  \label{ComplexQ22Matrix}
\end{equation}%
where $\sigma _{1},$ $\sigma _{2},$ $\sigma _{3}$ are the standard Pauli
matrices. The corresponding transformation law,%
\begin{equation}
Q^{\prime }\left( x^{\prime }\right) =S_{\Lambda }Q\left( x\right)
S_{\Lambda }^{-1},\qquad x^{\prime }=\Lambda x,\qquad S_{\Lambda }=\exp
\left( \frac{1}{4}\theta _{\mu \nu }\Sigma ^{\mu \nu }\right) ,
\label{ComplexQ22Transform}
\end{equation}%
under a proper Lorentz transformation preserves two invariants $\mathrm{tr}%
Q=0$ and $\det Q=-\mathbf{F}^{2}.$ Here, $S_{\Lambda }\sigma _{p}S_{\Lambda
}^{-1}=a_{pq}\sigma _{q}$ and $F_{p}^{\prime }\left( x^{\prime }\right)
=a_{pq}F_{q}\left( x\right) ,$ with $a_{qr}a_{qs}=\delta _{rs}$ for a given
complex orthogonal $3\times 3$ matrix (see section~5.1).

In this \textquotedblleft spinor\textquotedblright\ representation, one gets%
\begin{equation}
M^{\alpha\beta}Q=-\left. \left( \frac{d}{d\theta_{\alpha\beta}}Q^{\prime
}\left( \Lambda x\right) \right) \right\vert _{\theta_{\alpha\beta}=0}=-%
\frac{1}{2}\left( \Sigma^{\alpha\beta}Q-Q\Sigma^{\alpha\beta}\right)
\label{MaxwellSpinorGenerators}
\end{equation}
for generators of the one-parameter subgroups. Here, as in the case of
Weyl's equation, the matrices $\Sigma^{\alpha\beta}$ are given by (\ref%
{2generators}), but now, in view of (\ref{MaxwellSpinorGenerators}),
equation (\ref{WrongHelicity}) with $\lambda=-1$ takes the form,%
\begin{equation}
\frac{1}{2}\left( \Sigma_{\mu\nu}\partial^{\nu}Q-\partial^{\nu}Q\Sigma
_{\mu\nu}\right) =\partial_{\mu}Q,  \label{MaxwellSpinor}
\end{equation}
when the self-duality property (\ref{2SpinorSelfDuality}) is applied.

Equations (\ref{MaxwellSpinor}), obtained with the aid of the Pauli-Luba\'{n}%
ski vector, are equivalent to the system of complex Maxwell equations in
vacuum,%
\begin{equation}
\func{div}\mathbf{F}=0,\qquad\func{curl}\mathbf{F}=i\partial_{0}\mathbf{F}.
\label{MaxwellF}
\end{equation}
Indeed, when $\mu=0,$ with the help of (\ref{2generatorsdown}) one gets%
\begin{equation*}
-\frac{1}{2}\left( \sigma_{p}\partial_{p}Q-\partial_{p}Q\sigma_{p}\right)
=\partial_{0}Q,
\end{equation*}
or%
\begin{equation*}
-\left( \sigma_{p}\sigma_{q}-\sigma_{q}\sigma_{p}\right)
\partial_{p}F_{q}=2\partial_{0}\left( \sigma_{r}F_{r}\right) ,
\end{equation*}
which gives the second complex Maxwell equation (\ref{MaxwellF}) in view of
the commutation relation, $\left[ \sigma_{p},\sigma_{q}\right]
=2ie_{pqr}\sigma_{r}.$

When $\mu =p=1,2,3,$ in a similar fashion,%
\begin{equation}
2\partial _{p}Q=\partial _{0}Q\sigma _{p}-\sigma _{p}\partial
_{0}Q+ie_{pqr}\left( \partial _{q}Q\sigma _{r}-\sigma _{r}\partial
_{q}Q\right) ,
\end{equation}%
and letting $Q=F_{s}\sigma _{s},$ we obtain%
\begin{align*}
2\partial _{p}\left( F_{s}\sigma _{s}\right) & =\left( \sigma _{s}\sigma
_{p}-\sigma _{p}\sigma _{s}\right) \partial _{0}F_{s} \\
& +ie_{pqr}\left( \sigma _{s}\sigma _{r}-\sigma _{r}\sigma _{s}\right)
\partial _{q}F_{s}.
\end{align*}%
\newline
Evaluation of the commutators,%
\begin{equation}
\sigma _{s}\partial _{p}F_{s}=ie_{spl}\sigma _{l}\partial
_{0}F_{s}+e_{pqr}e_{slr}\sigma _{l}\partial _{q}F_{s},
\end{equation}%
and a familiar identity,%
\begin{equation}
e_{pqr}e_{slr}=\left\vert
\begin{array}{cc}
\delta _{ps} & \delta _{pl} \\
\delta _{qs} & \delta _{ql}%
\end{array}%
\right\vert =\delta _{ps}\delta _{ql}-\delta _{pl}\delta _{qs},
\end{equation}%
result in the system of Maxwell's equations (\ref{MaxwellF}).

\subsection{Traditional spinor form of Maxwell's equations}

Equations (\ref{MaxwellSpinor}), that are obtained here with the help of the
Pauli-Luba\'{n}ski vector, give an alternative (vacuum) version of the
spinor form of Maxwell's equations,%
\begin{equation}
\left( \partial _{0}+\boldsymbol{\sigma }\cdot \nabla \right) \left(
\boldsymbol{\sigma }\cdot \mathbf{F}\right) =j_{0}+\boldsymbol{\sigma }\cdot
\mathbf{j},  \label{Quanterions}
\end{equation}%
or, explicitly,%
\begin{equation}
\left(
\begin{array}{cc}
\partial _{0}+\partial _{3} & \partial _{1}-i\partial _{2}\medskip \\
\partial _{1}+i\partial _{2} & \partial _{0}-\partial _{3}%
\end{array}%
\right) \left(
\begin{array}{cc}
F_{3} & F_{1}-iF_{2}\medskip \\
F_{1}+iF_{2} & -F_{3}%
\end{array}%
\right) =\left(
\begin{array}{cc}
j_{0}+j_{3} & j_{1}-ij_{2}\medskip \\
j_{1}+ij_{2} & j_{0}-j_{3}%
\end{array}%
\right) ,  \label{QuanterionsTwo}
\end{equation}%
which was originally established in \cite{LapUh31} (see also \cite%
{RumerFetQFT}).

Let $Q=\boldsymbol{\sigma }\cdot \mathbf{F}$ and $\mathcal{D}=\sigma ^{\mu
}\partial _{\mu },$ $\mathcal{J}=\sigma ^{\mu }j_{\mu },$ when $\mathcal{D}Q=%
\mathcal{J}.$ Then $S_{\Lambda }QS_{\Lambda ^{-1}}=Q^{\prime }$ and, in view
of (\ref{LorentzOrthogonality}) and (\ref{SigmaAlternative}), one gets%
\begin{equation}
S_{\Lambda ^{-1}}^{\dag }\mathcal{D}S_{\Lambda ^{-1}}=\mathcal{D}^{\prime
},\qquad S_{\Lambda ^{-1}}^{\dag }\mathcal{J}S_{\Lambda ^{-1}}=\mathcal{J}%
^{\prime }
\end{equation}%
under a proper Lorentz transformation. In \textquotedblleft
new\textquotedblright\ coordinates, equation (\ref{QuanterionsTwo}) should
take a compact form, $\mathcal{D}^{\prime }Q^{\prime }=J^{\prime }.$ Thus%
\begin{equation}
\left( S_{\Lambda ^{-1}}^{\dag }\mathcal{D}S_{\Lambda ^{-1}}\right)
S_{\Lambda }QS_{\Lambda ^{-1}}=S_{\Lambda ^{-1}}^{\dag }\mathcal{J}%
S_{\Lambda ^{-1}},
\end{equation}%
or, $S_{\Lambda ^{-1}}^{\dag }\left( \mathcal{D}Q=\mathcal{J}\right)
S_{\Lambda ^{-1}},$ as a short proof of the covariance of Maxwell's
equations.

\section{Massive symmetric four-tensor field}

\subsection{Group-theoretical derivation}

The relativistic wave equation for a massive particle of spin two, described
by a real or complex symmetric four-tensor field $A^{\mu \nu }\left(
x\right) =A^{\nu \mu }\left( x\right) $ (see \cite{Fierz39}, \cite%
{FierzPauli39a}, \cite{FierzPauli39b}), can be obtained in a way that is
similar to our study of the Proca equation in section~4. Once again,%
\begin{eqnarray}
w_{\mu }A^{\alpha \beta } &=&\frac{1}{2}e_{\mu \nu \sigma \tau }\partial
^{\nu }\left( M^{\sigma \tau }A^{\alpha \beta }\right) \\
&=&-g^{\alpha \nu }e_{\mu \nu \sigma \tau }\partial ^{\sigma }A^{\tau \beta
}-g^{\beta \nu }e_{\mu \nu \sigma \tau }\partial ^{\sigma }A^{\tau \alpha }
\notag
\end{eqnarray}%
and%
\begin{equation}
e_{\mu \nu \sigma \tau }\partial ^{\sigma }A^{\tau \beta }=\frac{1}{2}e_{\mu
\nu \sigma \tau }F^{\sigma \tau \beta }=-G_{\mu \nu }^{\quad \beta },
\end{equation}%
where, by definition, $F^{\sigma \tau \beta }=\partial ^{\sigma }A^{\tau
\beta }-\partial ^{\tau }A^{\sigma \beta }$ and $e_{\mu \nu \sigma \tau
}F^{\sigma \tau \beta }=-2G_{\mu \nu }^{\quad \beta }.$ Thus%
\begin{equation}
w_{\mu }A^{\alpha \beta }=g^{\alpha \nu }G_{\mu \nu }^{\quad \beta
}+g^{\beta \nu }G_{\mu \nu }^{\quad \alpha },
\end{equation}%
or%
\begin{equation}
w^{\mu }A^{\alpha \beta }=G^{\mu \alpha \beta }+G^{\mu \beta \alpha },\qquad
w_{\mu }A_{\alpha \beta }=G_{\mu \alpha \beta }+G_{\mu \beta \alpha }
\label{WFP1}
\end{equation}%
and%
\begin{equation}
w^{2}A^{\alpha \beta }=w_{\mu }\left( w^{\mu }A^{\alpha \beta }\right)
=w_{\mu }G^{\mu \alpha \beta }+w_{\mu }G^{\mu \beta \alpha }.  \label{WFP2}
\end{equation}

In a similar fashion, one can show that%
\begin{eqnarray}
w_{\mu }G^{\mu \alpha \beta } &=&2\partial _{\nu }F^{\nu \alpha \beta
}+\partial _{\nu }F^{\nu \beta \alpha }  \label{FPGIdentity} \\
&&+\partial ^{\alpha }\left( F^{\beta \sigma \tau }g_{\sigma \tau }\right)
-g^{\alpha \beta }\partial _{\nu }\left( F^{\nu \sigma \tau }g_{\sigma \tau
}\right) ,  \notag
\end{eqnarray}%
as a result of an elementary but rather tedious four-tensor algebra
calculation with the help of a familiar relation%
\begin{equation}
e^{\mu \nu \sigma \tau }e_{\mu \kappa \lambda \rho }=-\left\vert
\begin{array}{ccc}
\delta _{\kappa }^{\nu } & \delta _{\lambda }^{\nu } & \delta _{\rho }^{\nu }
\\
\delta _{\kappa }^{\sigma } & \delta _{\lambda }^{\sigma } & \delta _{\rho
}^{\sigma } \\
\delta _{\kappa }^{\tau } & \delta _{\lambda }^{\tau } & \delta _{\rho
}^{\tau }%
\end{array}%
\right\vert .
\end{equation}

For a massive spin-2 field, the second Casimir operator is equal to $%
w^{2}A^{\alpha \beta }=-6m^{2}A^{\alpha \beta },$ and we obtain%
\begin{eqnarray}
&&\partial _{\nu }F^{\nu \alpha \beta }+\partial _{\nu }F^{\nu \beta \alpha
}-\frac{2}{3}g^{\alpha \beta }\partial _{\nu }\left( F^{\nu \sigma \tau
}g_{\sigma \tau }\right)  \label{FierzPauliFA} \\
&&\quad +\frac{1}{3}\partial ^{\alpha }\left( F^{\beta \sigma \tau
}g_{\sigma \tau }\right) +\frac{1}{3}\partial ^{\beta }\left( F^{\alpha
\sigma \tau }g_{\sigma \tau }\right) =-2m^{2}A^{\alpha \beta },  \notag
\end{eqnarray}%
where%
\begin{eqnarray}
F^{\nu \alpha \beta } &=&\partial ^{\nu }A^{\alpha \beta }-\partial ^{\alpha
}A^{\nu \beta }=-F^{\alpha \nu \beta },  \label{FtensorFP} \\
F^{\nu \sigma \tau }g_{\sigma \tau } &=&\partial ^{\nu }A-\partial _{\tau
}A^{\nu \tau },\qquad A=g_{\sigma \tau }A^{\sigma \tau },  \notag
\end{eqnarray}%
and $F^{\alpha \beta \gamma }+F^{\beta \gamma \alpha }+F^{\gamma \alpha
\beta }=0$ provided $A^{\alpha \beta }=A^{\beta \alpha }.$

In terms of potentials, one gets%
\begin{eqnarray}
&&\partial ^{2}A^{\alpha \beta }-\frac{2}{3}\left( \partial ^{\alpha
}\partial _{\nu }A^{\nu \beta }+\partial ^{\beta }\partial _{\nu }A^{\nu
\alpha }\right) +\frac{1}{3}\partial ^{\alpha }\partial ^{\beta }A
\label{FierzPauliA} \\
&&\quad -\frac{1}{3}g^{\alpha \beta }\left( \partial ^{2}A-\partial _{\mu
}\partial _{\nu }A^{\mu \nu }\right) =-m^{2}A^{\alpha \beta },  \notag
\end{eqnarray}%
with $\partial _{\mu }\partial _{\nu }A^{\mu \nu }=0,$ if $m>0,$ in view of%
\begin{equation*}
0\equiv w^{2}\left( \partial _{\mu }\partial _{\nu }A^{\mu \nu }\right)
=\partial _{\mu }\partial _{\nu }\left( w^{2}A^{\mu \nu }\right)
=-6m^{2}\partial _{\mu }\partial _{\nu }A^{\mu \nu }.
\end{equation*}%
The same results can be derived by differentiation from (\ref{FierzPauliFA})
or, independently, with the help of an operator identity%
\begin{equation*}
w^{2}=-\frac{1}{2}\partial ^{2}\left( M^{\sigma \tau }M_{\sigma \tau
}\right) -\partial _{\mu }\partial ^{\nu }\left( M^{\mu \sigma }M_{\sigma
\nu }\right) .
\end{equation*}%
Moreover, equation (\ref{FierzPauliA}) is reduced to%
\begin{equation}
2\left( \partial ^{\alpha }\partial _{\nu }A^{\nu \beta }+\partial ^{\beta
}\partial _{\nu }A^{\nu \alpha }\right) -\partial ^{\alpha }\partial ^{\beta
}A+g^{\alpha \beta }\partial ^{2}A=0  \label{Gcontraction}
\end{equation}%
with the help of the first Casimir operator, $\partial ^{2}A^{\alpha \beta
}=-m^{2}A^{\alpha \beta }.$ Multiplication of (\ref{Gcontraction}) by $%
g_{\alpha \beta }$ with contraction over two repeated indices results in $%
\partial ^{2}A=0.$ Then $A=0,$ due to%
\begin{equation*}
0=\partial ^{2}A=\partial ^{2}\left( g_{\mu \nu }A^{\mu \nu }\right) =g_{\mu
\nu }\partial ^{2}\left( A^{\mu \nu }\right) =-m^{2}A,\qquad m>0.
\end{equation*}

Thus, the system of wave equations for a massive spin-2 particle has the form%
\begin{equation}
\partial ^{2}A^{\mu \nu }+m^{2}A^{\mu \nu }=0,\qquad \partial ^{\alpha
}\left( \partial _{\nu }A^{\nu \beta }\right) +\partial ^{\beta }\left(
\partial _{\nu }A^{\nu \alpha }\right) =0,  \label{FierzPauliFinal}
\end{equation}%
subject to $A=g_{\mu \nu }A^{\mu \nu }=\partial _{\mu }\partial _{\nu
}A^{\mu \nu }=0.$ It is worth noting, once again, that we have derived these
equations by using the Pauli-Luba\'{n}ski vector and the relativistic
definition of mass and spin in terms of Casimir operators of the Poincar\'{e}
group.

In the massless limit $m\rightarrow 0,$ instead of (\ref{FierzPauliFinal}),
one has $\partial ^{2}A^{\mu \nu }=0$ and $\partial ^{2}A=0$ subject to%
\begin{equation}
\partial ^{\alpha }\left( \partial _{\nu }A^{\nu \beta }-\frac{1}{4}\partial
^{\beta }A\right) +\partial ^{\beta }\left( \partial _{\nu }A^{\nu \alpha }-%
\frac{1}{4}\partial ^{\alpha }A\right) =\frac{1}{2}g^{\alpha \beta }\partial
_{\mu }\partial _{\nu }A^{\mu \nu },
\end{equation}%
in a similar fashion. Moreover, if $m=0,$ equation (\ref{FierzPauliFA}) is
invariant under a familiar gauge transformation $A_{\alpha \beta
}\rightarrow A_{\alpha \beta }^{\prime }+\partial _{\alpha }f_{\beta
}+\partial _{\beta }f_{\alpha }$ provided that $\partial ^{\nu }\left(
\partial _{\nu }f_{\alpha }-\partial _{\alpha }f_{\nu }\right) =0$
(Maxwell's equations in vacuum).

\subsection{An alternative gauge condition}

In view of the following identity,%
\begin{eqnarray*}
&&\partial ^{\alpha }\left( \partial _{\nu }A^{\nu \beta }-\frac{1}{4}%
\partial ^{\beta }A\right) +\partial ^{\beta }\left( \partial _{\nu }A^{\nu
\alpha }-\frac{1}{4}\partial ^{\alpha }A\right) \\
&&\quad =\partial ^{\alpha }\partial _{\nu }A^{\nu \beta }+\partial ^{\beta
}\partial _{\nu }A^{\nu \alpha }-\frac{1}{2}\partial ^{\alpha }\partial
^{\beta }A,
\end{eqnarray*}%
one can impose another condition,%
\begin{equation}
4\partial _{\nu }A^{\mu \nu }-\partial ^{\mu }A=0,  \label{4gaugeFP}
\end{equation}%
in order to simplify (\ref{FierzPauliA}):%
\begin{equation}
\partial ^{2}A^{\mu \nu }-\frac{1}{4}g^{\mu \nu }\partial ^{2}A=-m^{2}A^{\mu
\nu }.
\end{equation}%
Moreover, by contraction,\footnote{%
Condition $A=0$ is not required in the massless limit.}%
\begin{equation*}
0\equiv \partial ^{2}A-\frac{1}{4}g_{\mu \nu }g^{\mu \nu }\partial
^{2}A=-m^{2}A,\qquad A=g_{\sigma \tau }A^{\sigma \tau }=0,
\end{equation*}%
if $m>0.$ As a result, we obtain%
\begin{equation}
\partial ^{2}A^{\mu \nu }+m^{2}A^{\mu \nu }=0,\qquad \partial _{\nu }A^{\mu
\nu }=\partial _{\nu }A^{\nu \mu }=0.  \label{FierzPauliAGauge}
\end{equation}%
These equations were originally introduced by Fierz and Pauli \cite{Fierz39}%
, \cite{FierzPauli39a}, \cite{FierzPauli39b} with the help of a Lagrangian
approach. It is worth noting that our equations (\ref{FierzPauliFinal}) are
necessary and sufficient with the relativistic definition of mass and spin-2
of the field in question, whereas traditional equations (\ref%
{FierzPauliAGauge}) give only sufficient conditions.

\subsection{Fierz-Pauli vs Maxwell's equations}

When $m=0,$ one gets%
\begin{equation}
\partial _{\nu }F^{\mu \nu \alpha }=\partial _{\nu }\left( \partial ^{\mu
}A^{\nu \alpha }-\partial ^{\nu }A^{\mu \alpha }\right) =\partial ^{\mu
}\left( \partial _{\nu }A^{\nu \alpha }\right) -\partial ^{2}A^{\mu \alpha
}=0,
\end{equation}%
subject to (\ref{FierzPauliAGauge}). In addition,%
\begin{equation}
\partial _{\lambda }F_{\sigma \tau \alpha }+\partial _{\sigma }F_{\tau
\lambda \alpha }+\partial _{\tau }F_{\lambda \sigma \alpha }=0,
\end{equation}%
which follows from definition. These facts allow one to represent the
massless Fierz-Pauli equations in terms of the third rank field tensor,
somewhat similar to classical electrodynamics. Indeed, by analogy with
Maxwell's equations, we obtain%
\begin{equation}
\partial _{\nu }F^{\mu \nu \alpha }=0,\qquad \partial _{\nu }G^{\mu \nu
\alpha }=0  \label{MaxwellFierzPauli}
\end{equation}%
in view of $2G^{\mu \nu \alpha }=-e^{\mu \nu \sigma \tau }F_{\sigma \tau
}^{\quad \alpha }$ and
\begin{equation*}
\partial _{\nu }G^{\mu \nu \alpha }=-\frac{1}{6}e^{\mu \nu \sigma \tau
}\left( \partial _{\nu }F_{\sigma \tau }^{\quad \alpha }+\partial _{\sigma
}F_{\tau \nu }^{\quad \alpha }+\partial _{\tau }F_{\nu \sigma }^{\quad
\alpha }\right) =0
\end{equation*}%
(for every fixed $\alpha =0,1,2,3).$

Finally, both pairs of these equations can be combined together in the
following complex form%
\begin{equation}
\partial _{\nu }Q^{\mu \nu \alpha }=0,\qquad Q^{\mu \nu \alpha }=F^{\mu \nu
\alpha }-\frac{i}{2}e^{\mu \nu \sigma \tau }F_{\sigma \tau }^{\quad \alpha },
\label{FierzPauliMaxwell}
\end{equation}%
with the help of a self-dual complex four-tensor:%
\begin{equation}
2iQ^{\mu \nu \alpha }=e^{\mu \nu \sigma \tau }Q_{\sigma \tau }^{\quad \alpha
},\quad e_{\mu \nu \sigma \tau }Q^{\sigma \tau \alpha }=2iQ_{\mu \nu
}^{\quad \alpha }.  \label{MaxwellFierzPauliDual}
\end{equation}%
The covariant field \textquotedblleft energy-momentum\textquotedblright\
tensor and the corresponding differential balance equation,%
\begin{equation}
\frac{\partial }{\partial x^{\nu }}\left( Q_{\mu \lambda \sigma }^{\ast
}Q^{\lambda \nu \sigma }+Q_{\mu \lambda \sigma }\overset{\ast }{\left.
Q^{\lambda \nu \sigma }\right. }\right) =0,  \label{EnergyMomentumBalance}
\end{equation}%
can be derived in a complete analogy with complex electrodynamics \cite%
{KrLanSus150}, \cite{KrLanSu15}.

\subsection{Fierz-Pauli vs linearized Einstein's equations}

In general relativity, the linearized equations for a weak gravitational
field \cite{Einstein}, \cite{Hilbert}, namely,%
\begin{eqnarray}
&&\partial _{\mu }\partial ^{\sigma }h_{\sigma \nu }+\partial _{\nu
}\partial ^{\sigma }h_{\sigma \mu }-\partial _{\mu \nu }h
\label{EinsteinLin} \\
&&\quad -\partial ^{2}h_{\mu \nu }-g_{\mu \nu }\left( \partial _{\sigma
}\partial _{\tau }h^{\sigma \tau }-\partial ^{2}h\right) =0,  \notag
\end{eqnarray}%
describe small deviations from the flat Minkowski metric, $g_{\mu \nu }=%
\mathrm{diag}\left( 1,-1,-1,-1\right) ,$ on the pseudo-Riemannian manifold
subject to a gauge condition%
\begin{equation}
2\partial ^{\nu }h_{\mu \nu }-\partial _{\mu }h=0,\qquad h=g^{\sigma \tau
}h_{\sigma \tau }  \label{EinsteinLinGauge}
\end{equation}%
(see, for example, \cite{Dam07}, \cite{Faddeev82}, \cite{FierzPauli39b},
\cite{Fock64}, \cite{Iorio15}, \cite{Khriplovich}, \cite{LanLif2}, \cite%
{MisThoWhe}, \cite{ParkTau82}, \cite{VizginSmorod}, \cite{Wald84}, \cite%
{Wein72}, \cite{WeinCosm}, \cite{Weyl44}, \cite{Witten81}, \cite{Zakharov73}%
, and the references therein for more details).

Our calculations have shown that linearized Einstein's equations (\ref%
{EinsteinLin}) do not coincide with the massless limit of the spin-2
particle wave equation (\ref{FierzPauliA}). But they can be reduced to the
massless case of the Fierz-Pauli equations (\ref{FierzPauliAGauge}) in view
of an additional condition (\ref{EinsteinLinGauge}) on a certain solution
set. In the literature, this fact is usually interpreted as spin-2 for a
graviton although, from the group-theoretical point of view, this massless
limit yet requires certain analysis of helicity, say similar to the one in
electrodynamics \cite{KrLanSus150}, which will be discuss elsewhere.

\section{Summary}

In this article, we analyze kinematics of the fundamental relativistic wave
equations, in a traditional way, from the viewpoint of the representation
theory of the Poincar\'{e} group. In particular, the importance of the
Pauli-Luba\'{n}ski pseudo-vector is emphasized here not only for the
covariant definition of spin and helicity of a given field but also for the
derivation of the corresponding equation of motion from first principles. In
this consistent group-theoretical approach, the resulting wave equations
occur, in general, in certain overdetermined forms, which can be reduced to
the standard ones by a matrix version of Gaussian elimination.

Although, mathematically, all representations of the Poincar\'{e} group are
locally equivalent \cite{BargmannWigner48}, their explicit realizations in
conventional linear spaces of four-vectors and tensors, spinors and
bispinors, etc. are quite different from the viewpoint of physics. This is
why, as the reader can see in the table below, the corresponding
relativistic wave equations are so different.\bigskip

\begin{tabular}{|l|l|l|}
\hline
\textit{Classical field} & \textit{Transformation law (a law of nature)} &
\textit{Wave equation} \\ \hline
Bispinor (\ref{bispinor}) & $\psi ^{\prime }\left( x^{\prime }\right)
=S_{\Lambda }\psi \left( x\right) ,\ x^{\prime }=\Lambda x;\ $see (\ref%
{MatrixS}) and (\ref{OneParameter}) & Dirac~(\ref{DiracEquationGamma}) \\
\hline
Spinor (\ref{2spinor}) & $\psi ^{\prime }\left( x^{\prime }\right)
=S_{\Lambda }\psi \left( x\right) ,\ x^{\prime }=\Lambda x;\ $see (\ref%
{SpinorTransform2}) and (\ref{OneParameter}) & Weyl (\ref{Weyl2}) \\ \hline
Four-vector & $A^{\prime \mu }\left( x^{\prime }\right) =\Lambda _{\ \nu
}^{\mu }A^{\nu }\left( x\right) ;\ $see (\ref{OneParameter}) & Proca (\ref%
{ProcaEquation}) \\ \hline
\textquotedblleft Feynman slash\textquotedblright & $Q^{\prime }\left(
x^{\prime }\right) =S_{\Lambda }Q\left( x\right) S_{\Lambda }^{-1},\
x^{\prime }=\Lambda x;\ $see (\ref{ProcaBispinor2}) & Proca (\ref%
{ProcaEquation}) \\ \hline
Four-tensor (\ref{4TensorQ}) & $Q^{\prime \mu \nu }\left( x^{\prime }\right)
=\Lambda _{\ \sigma }^{\mu }\Lambda _{\ \tau }^{\nu }Q^{\sigma \tau }\left(
x\right) ,\ x^{\prime }=\Lambda x;\ $see (\ref{OneParameter}) & Maxwell (\ref%
{4TensorQ}, \ref{MaxwellQuaterion}) \\ \hline
Complex $3D$ vector & $\mathbf{F}^{\prime }\left( x^{\prime }\right)
=S_{\Lambda }\mathbf{F}\left( x\right) ,$ $x^{\prime }=\Lambda x;\ $see
Sect.~5.1, (\ref{OneParameter}) & Maxwell (\ref{MaxwellF}) \\ \hline
Complex matrix (\ref{ComplexQ22Matrix}) & $Q^{\prime }\left( x^{\prime
}\right) =S_{\Lambda }Q\left( x\right) S_{\Lambda }^{-1},\ x^{\prime
}=\Lambda x;\ $see (\ref{ComplexQ22Transform}) & Maxwell (\ref{MaxwellF}) \\
\hline
Symmetric four-tensor & $A^{\prime \mu \nu }\left( x^{\prime }\right)
=\Lambda _{\ \sigma }^{\mu }\Lambda _{\ \tau }^{\nu }A^{\sigma \tau }\left(
x\right) ,\ x^{\prime }=\Lambda x$ & Fierz-Pauli (\ref{FierzPauliFinal}) \\
\hline
\end{tabular}
\bigskip

\noindent {\textbf{Acknowledgements}}. This research was partially supported
by NSF grants DMS~\#~1535833 and DMS~\#~1151618. We are indebted to Prof.
Dr. Albert Boggess, Prof. Dr. Mark~P Faifman, Prof. Dr. Gerald A. Goldin,
Prof. Dr.~John Klauder, Prof. Dr. Margarita A. Man'ko, Prof. Dr. Vladimir I.
Man'ko, Prof. Dr. Svetlana Roudenko, and Prof. Dr. Igor N. Toptygin for
their valuable comments and encouragement. Comments from the referees, which
have helped to improve the presentation, are much appreciated.


\begin{thebibliography}{99}
\bibitem{Akh:Ber} Akhiezer~A and Berestetskii~V~B 1965 \textit{Quantum
Electrodynamics\/} (New York: Interscience).

\bibitem{Bargmann47} Bargmann~V~1947 {Irreducible unitary representations of
the Lorentz group\/} \textit{Ann. Math.\/} \textbf{48} 568--640.

\bibitem{Bargmann54} Bargmann~V~1954 {On unitary ray representations of
continuous groups\/} \textit{Ann. Math.\/} \textbf{59} 1--46.

\bibitem{BargmannWigner48} Bargmann~V and Wigner~E~1948 {Group theoretical
discussion of relativistic wave equations\/} \textit{Proc. Nat. Acad. Sci.
U.S.A.\/} \textbf{34} 211--223.

\bibitem{BarutRaczka80} Barut~A~O and Raczka~R~1986 \textit{Theory of
Group Representations and Applications\/} 2nd edn (Singapore: World
Scientific).

\bibitem{Ber:Lif:Pit} Berestetskii~V~B, Lifshitz~E~M and Pitaevskii~L~P~1971
\textit{Relativistic Quantum Theory\/} (Oxford: Pergamon).

\bibitem{Bhabha45} Bhabha~H~J~1945 {Relativistic wave equations for the
elementary particles\/} \textit{Rev.~Mod.~Phys.\/} \textbf{17} 200--216.

\bibitem{Bia:Bia75} Bia\l ynicki-Birula~I and Bia\l ynicki-Birula~Z~1975
\textit{Quantum Electrodynamics\/} (Oxford, New York: Pergamon and
PWN--Polish Scientific Publishers).

\bibitem{BogolubovShirkov} Bogoliubov~N~N and Shirkov~D~V~1980 \textit{%
Introduction to the Theory of Quantized Fields\/} 3rd edn (New York: Wiley).

\bibitem{Bogolubovetal90} Bogolubov~N~N, Logunov~A~A, Oksak~A~I and
Todorov~I~T~1990 \textit{General Principles of Quantum Field Theory\/}
(Dordrecht: Kluwer).

\bibitem{Dam07} Damour~T~2007 {General relativity today\/} \textit{%
Gravitation and Experiment - Poincar\'{e} Seminar 2006\/} ed T~Damour,
B~Duplantier and V~Rivasseau (Basel: Birkh\"{a}user) pp~1--49.

\bibitem{Dirac28} Dirac~P~A~M~1928 {The quantum theory of the electron\/}
\textit{Proc. Roy. Soc. London\/} \textbf{A117} 610--624.

\bibitem{Dirac36} Dirac~P~A~M~1936 {Relativistic wave equations\/} \textit{%
Proc. Roy. Soc. London\/} \textbf{A155} 447--459.

\bibitem{Einstein} Einstein~A~1916 {N\"{a}herungsweise Integration der
Feldgleichungen der Gravitation\/} \textit{Sitzungsber. K\"{o}n. Preu\ss .
Akad. Wiss. zu Berlin }\textbf{1} 688--696.

\bibitem{Faddeev82} Faddeev~L~D~1982 {The energy problem in Einstein's
theory of gravitation\/} \textit{Phys. Usp. }\textbf{25} 130--142.

\bibitem{Feynman} Feynman~R~P~1998 \textit{The Theory of Fundamental
Processes\/} (Cambridge: Perseus).

\bibitem{Fierz39} Fierz~M~1939 {\"{U}ber die relativistische Theorie kr\"{a}%
ftefreier Teilchen mit beliebigem Spin\/} \textit{Helv. Phys.~Acta\/}
\textbf{12} 3--37.

\bibitem{Fierz40} Fierz~M~1940 {\"{U}ber den Drehimpuls von Teilichen mit
Ruhemasse null und beliebigem Spin\/} \textit{Helv. Phys.~Acta\/} \textbf{13}
45--60.

\bibitem{FierzPauli39a} Fierz~M~and Pauli~W~1939 {\"{U}ber die
relativistische Fieldgleichungen von Teilchen mit beliebigem Spin in
elektromagnetischen Feld\/} \textit{Helv. Phys.~Acta\/} \textbf{12} 297--300.

\bibitem{FierzPauli39b} Fierz~M~and Pauli~W~1939 {On relativistic wave
equations for particles of arbitrary spin in an electromagnetic field\/}
\textit{Proc. Roy. Soc. London\/} \textbf{A173} 211--232.

\bibitem{Fock64} Fock~V~A~1965 \textit{The Theory of Space, Time and
Gravitation\/} (New York: Pergamon).

\bibitem{Gantmacher1} Gantmacher~F~R~2000 \textit{The Theory of Matrices\/}
\textit{\/}vol~1 (Providence, Rhode Island: AMS Chelsea).

\bibitem{Gelfandetal} Gel'fand~I~M, Minlos~R~A and Shapiro~Z~Ya~1963 \textit{%
Representations of the Rotation and Lorentz Groups and their Applications\/}
(New York: Pergamon).

\bibitem{Greiner} Greiner~W 2000 \textit{Relativistic Quantum Mechanics:
Wave Equations\/} 3rd edn (Berlin, Heidelberg, New York: Springer).

\bibitem{Hilbert} Hilbert~D~1924 {Die Grundlagen der Physik (Zweite
Mitteilung)\/} \textit{Math. Ann. }\textbf{92} 1--32.

\bibitem{Iorio15} Iorio~L~2015 {Editorial for the special issue 100 years of
chronogeometrodynamics: the status of the Einstein's theory of gravitation
in its cetennial year\/} \textit{Universe\/} \textbf{1} 38--81.

\bibitem{ItzZub} Itzykson~C and Zuber~J-B~2005 \textit{Quantum Field Theory\/%
} (New York: Dover).

\bibitem{Khriplovich} Khriplovich~I~B~2005 \textit{General Relativity\/}
(New York: Springer).

\bibitem{Kajita15} Kajita~T~2015 {\/Discovery of atmospheric neotrino
oscillations\/} \textit{Nobel Lecture\/}, December 8;
http://www.nobelprize.org/nobel%
\_prizes/physics/laureates/2015/kajita-lecture.html

\bibitem{Klyshko94} Klyshko~D.~N.~1994 {Quantum optics: quantum, classical,
and metaphysical aspects\/} \textit{Phys.-Usp.\/} \textbf{37} 1097--1123;
see also: \textit{Annals of the New York Academy of Sciences\/} vol.~755
\textit{Fundamental Problems in Quantum Theory\/} April~1995 pp~13--27.

\bibitem{Kramersetal41} Kramers~H~A, Belinfante~F~J and Luba\'{n}%
ski~J~K~1941 {\"{U}ber freie Teilchen mit nichtverschwindender Masse und
beliebiger Spinquantenzahl\/} \textit{Physica\/} \textbf{8} 597--627.

\bibitem{KrLanSus150} Kryuchkov~S~I, Lanfear~N~A and Suslov~S~K~2015 {The
Pauli-Luba\'{n}ski vector, complex electrodynamics, and photon helicity }%
\textit{Physica Scripta\/} \textbf{90} 074065 (8pp)

\bibitem{KrLanSu15} Kryuchkov~S~I, Lanfear~N~A and Suslov~S~K~2015 {%
\/Complex Form of Classical and Quantum Electrodynamics\/} \textit{in
preparation\/}.

\bibitem{LanLif2} Landau~L~D~and Lifshitz~E~M 1975 \textit{The Classical
Theory of Fields}\textsl{\/} 4th edn (Oxford: Butterworth-Heinenann).

\bibitem{LapUh31} Laporte~O and Uhlenbeck~G~E~1931 {Applications of spinor
analysis to the Maxwell and Dirac equations\/} \textit{Phys. Rev.\/} \textbf{%
37} 1380--1397.

\bibitem{Lub41} Luba\'{n}ski~J~K~1941 {Sur le spin des particules \'{e}l\'{e}%
mentaires\/} \textit{Physica\/} \textbf{8} 44--52.

\bibitem{Lub42-I} Luba\'{n}ski~J~K~1942 {Sur la th\'{e}orie des particules
\'{e}l\'{e}mentaires de spin quelconque. I\/} \textit{Physica\/} \textbf{9}
310--324.

\bibitem{Lub42-II} Luba\'{n}ski~J~K~1942 {Sur la th\'{e}orie des particules
\'{e}l\'{e}mentaires de spin quelconque. II\/} \textit{Physica\/} \textbf{9}
325--338.

\bibitem{Majorana32} Majorana~E~1932 {Theoria relativistica di particelle
con momento intrinseco arbitrario\/} \textit{Nuovo Cimento\/} \textbf{9}
335--344; reprinted in English: {Relativistic theory of particles with
arbitrary intrinsic angular momentum\/}, Majorana~E~2006 \textit{Scientific
Papers\/} ed J~F~Bassani (Berlin, Heidelberg, New York: Springer)
pp~150--159.

\bibitem{McDonald15} McDonald~A~B~2015 {\/The Sunbury Neutrino Observatory:
Observation of flavor change for solar neotrinos\/} \textit{Nobel Lecture\/}%
, December 8;
http://www.nobelprize.org/nobel%
\_prizes/physics/laureates/2015/mcdonald-lecture.html

\bibitem{MinkowskiI} Minkowski~H~1908, {Die Grundlagen f\"{u}r die
elektromagnetischen Vorg\"{a}nge in bewegten K\"{o}rpern\/} \textit{Nachr. K%
\"{o}nig. Ges. Wiss. G\"{o}ttingen, math.-phys. Kl.\/} 53--111.

\bibitem{MisThoWhe} Misner~C~W, Thorne~K~S and Wheeler~J~A~1973 \textit{%
Gravitation\/} (San Francisco: Freeman).

\bibitem{MoskalevQFT} Moskalev~A~N~2006 \textit{Relativistic Field Theory\/}
(St. Peterburg: PIYaF RAN) (in Russian).

\bibitem{MukhWin07} Mukhanov~V and Winitzki~S~2007 \textit{Introduction to
Quantum Effects in Gravity\/} (Cambridge: Cambridge University Press).

\bibitem{Naber12} Naber~G~L~2012 \textit{The Geometry of Minkowski
Spacetime: An Introduction to the Mathematics of the Special Theory of
Relativity\/} 2nd edn (Berlin, Heidelberg, New York: Springer).

\bibitem{ParkTau82} Parker~E and Taubes~C~H~1982 {On Witten's proof of the
positive energy theorem\/} \textit{Comm. Math. Phys.~}\textbf{84} 223--238.

\bibitem{Pauli36} Pauli~W~1936 {Contributions math\'{e}matiques \`{a} la th%
\'{e}orie des matrices de Dirac\/} \textit{Ann.~Inst.~H.~Poincar\'{e}\/}
\textbf{6} 109--136.

\bibitem{Pauli41} Pauli~W~1941 {Relativistic field theories of elementary
particles\/} \textit{Rev.~Mod.~Phys.\/} \textbf{13} 203--232.

\bibitem{Pauli} Pauli~W~1958 \textit{Theory of Relativity\/} (Oxford:
Pergamon).

\bibitem{PeskinSchroeder} Peskin~M~E and Schroeder~D~V~1995 \textit{An
Introduction to Quantum Field Theory\/} (Boulder: Westview Press).

\bibitem{Pontecorvo83} Pontecorvo~B~M~1983 {Pages in the development of
neutrino physics\/} \textit{Phys. Usp. }\textbf{26} 1087--1108.

\bibitem{Proca36} Proca~A~1936 {Sur la th\'{e}orie ondulatoire des \'{e}%
lectrons positifs et n\'{e}gatifs\/} \textit{J.~Phys.~Radium\/} \textbf{7}
347--353.

\bibitem{Roy01} Roy~A~2001 {Discovery of parity violation: breakdown of a
symmetry principle\/} \textit{Resonance\/} \textbf{6} 32--43.

\bibitem{RumerFetQFT} Rumer~Yu~B and Fet~A~I~1977 \textit{Group Theory and
Quantized Fields\/} (Moscow: Nauka) (in Russian).

\bibitem{Ryder} Ryder~L~H~1996, \textit{Quantum Field Theory\/} 2nd edn
(Cambridge: Cambridge University Press).

\bibitem{ScheckQuantumPhysics} Scheck~F~2007 \textit{Quantum Physics\/}
(Berlin, Heidelberg, New York: Springer).

\bibitem{Schiff} Schiff~L~I~1968 \textit{Quantum Mechanics\/} \textit{\/}3rd
edn (New York: McGraw-Hill).

\bibitem{Schweber61} Schweber~S~S~1961 \textit{An Introduction to
Relativistic Quantum Field Theory\/} (Evanston, Illinois; Elmsford, New
York: Row, Peterson and Company).

\bibitem{SchweberBetheHoffmann} Schweber~S~S, Bethe~H~A and
Hoffmann~D~F~1955 \textit{Mesons and Fields\/} vol~1 (Evanston, Illinois;
Elmsford, New York: Row, Peterson and Company).

\bibitem{Sohnius85} Sohnius~M~F~1985 {Introducing supersymmetry\/} \textit{%
Phys.~Rep. }\textbf{128} 39--204.

\bibitem{ToptyginI} Toptygin~I~N~2014 \textit{Foundations of Classical and
Quantum Electrodynamics\/} (Weinheim: Wiley-VCH).

\bibitem{VMK} Varshalovich~D~A, Moskalev~A~N and Khersonskii~V~K~1988
\textit{Quantum Theory of Angular Momentum\/} (Singapore: World Scientific).

\bibitem{VizginSmorod} Vizgin~V~P and Smorodinski\u{\i}~Ya~A~1979 {From the
equivalence principle to the equations of gravitation\/} \textit{Phys. Usp. }%
\textbf{22} 489--513.

\bibitem{Wald84} Wald~R~N~1984 \textit{General Relativity\/} (Chicago:
University of Chicago Press).

\bibitem{Wein72} Weinberg~S~1972 \textit{Gravitation and Cosmology:
Principles and Applications of the General Theory of Relativity\/} (New
York: Wiley).

\bibitem{Wein} Weinberg~S~1998 \textit{The Quantum Theory of Fields\/} vol~1
(Cambridge: Cambridge University Press).

\bibitem{WeinCosm} Weinberg~S~2008 \textit{Cosmology\/} (Oxford: Oxford
University Press).

\bibitem{Weyl} Weyl~H~1929 {Gravitation and the electron\/} \textit{Proc.
Nat. Acad. Sci. U.S.A.\/} \textbf{15} 323--334.

\bibitem{Weyl44} Weyl~H~1944 {How far can one get with a linear field theory
of gravitation in flat space-time?\/} \textit{Amer. J. Math. }\textbf{66}
591--604.

\bibitem{Wigner39} Wigner~E~1939 {On unitary representations of the
inhomogeneous Lorentz group\/} \textit{Ann. Math.\/} \textbf{40} 149--204.

\bibitem{Witten81} Witten~E~1981 {A new proof of the positive energy
theorem\/} \textit{Comm. Math. Phys.~}\textbf{80} 381--402.

\bibitem{Wuetal57} Wu~C~S, Ambler~E, Hayward~R~W, Hoppes~D~D and
Hudson~R~P~1957 {Experimental test of parity conservation in beta decay\/}
\textit{Phys. Rev.\/} \textbf{105} 1413--1415.

\bibitem{Zakharov73} Zakharov~V~D~1973 \textit{Gravitational waves in
Einstein's theory\/} (New York: Wiley).
\end{thebibliography}
\end{document}